# Dangerous human-made interference with climate: A GISS modelE study

J. Hansen,[1,2] M. Sato,[2] R. Ruedy,[3] P. Kharecha,[2] A. Lacis,[1,4] R. Miller,[1,5] L. Nazarenko,[2] K. Lo,[3] G.A. Schmidt,[1,4] G. Russell,[1] I. Aleinov,[2] S. Bauer,[2] E. Baum,[6] B. Cairns,[5] V. Canuto,[1] M. Chandler,[2] Y. Cheng,[3] A. Cohen,[6] A. Del Genio,[1,4] G. Faluvegi,[2] E. Fleming,[7] A. Friend,[8] T. Hall,[1,5] C. Jackman,[7] J. Jonas,[2] M. Kelley,[8] N.Y. Kiang,[1] D. Koch,[2,9] G. Labow,[7] J. Lerner,[2] S. Menon,[10] T. Novakov,[10] V. Oinas,[3] Ja. Perlwitz,[5] Ju. Perlwitz,[2] D. Rind,[1,4] A. Romanou,[1,4] R. Schmunk,[3] D. Shindell,[1,4] P. Stone,[11] S. Sun,[1,11] D. Streets,[12] N. Tausnev,[3] D. Thresher,[4] N. Unger,[2] M. Yao,[3] S. Zhang[2]

We investigate the issue of "dangerous human-made interference with climate" using simulations with GISS modelE driven by measured or estimated forcings for 1880-2003 and extended to 2100 for IPCC greenhouse gas scenarios as well as the 'alternative' scenario of *Hansen and Sato* [2004]. Identification of 'dangerous' effects is partly subjective, but we find evidence that added global warming of more than 1ºC above the level in 2000 has effects that may be highly disruptive. The alternative scenario, with peak added forcing ~1.5 W/m$^2$ in 2100, keeps further global warming under 1°C if climate sensitivity is ~3ºC or less for doubled $CO_2$. The alternative scenario keeps mean regional seasonal warming within 2σ (standard deviations) of 20$^{th}$ century variability, but other scenarios yield regional changes of 5-10σ, i.e., mean conditions outside the range of local experience. We discuss three specific sub-global topics: Arctic climate change, tropical storm intensification, and ice sheet stability. We suggest that Arctic climate change has been driven as much by pollutants ($O_3$, its precursor $CH_4$, and soot) as by $CO_2$, offering hope that dual efforts to reduce pollutants and slow $CO_2$ growth could minimize Arctic change. Simulated recent ocean warming in the region of Atlantic hurricane formation is comparable to observations, suggesting that greenhouse gases (GHGs) may have contributed to a trend toward greater hurricane intensities. Increasing GHGs cause significant warming in our model in submarine regions of ice shelves and shallow methane hydrates, raising concern about the potential for accelerating sea level rise and future positive feedback from methane release. Growth of non-$CO_2$ forcings has slowed in recent years, but $CO_2$ emissions are now surging well above the alternative scenario. Prompt actions to slow $CO_2$ emissions and decrease non-$CO_2$ forcings are needed to achieve the low forcing of the alternative scenario.

[1]NASA Goddard Institute for Space Studies, New York, New York, USA.

[2]Columbia University Earth Institute, New York, New York, USA.

[3]Sigma Space Partners LLC, New York, New York, USA.

[4]Department of Earth and Environmental Sciences, Columbia University, New York, New York, USA.

[5]Department of Applied Physics and Applied Mathematics, Columbia University, New York, New York, USA.

[6]Clean Air Task Force, Boston, Massachusetts, USA.

[7]NASA Goddard Space Flight Center, Greenbelt, Maryland, USA.

[8]Laboratoire des Sciences du Climat et de l'Environnement, Orme des Merisiers, Gif-sur-Yvette Cedex, France.

[9]Department of Geology, Yale University, New Haven, Connecticut, USA.

[10]Lawrence Berkeley National Laboratory, Berkeley, California, USA.

[11]Massachusetts Institute of Technology, Cambridge, Massachusetts, USA.

[12]Argonne National Laboratory, Argonne, Illinois, USA.

## 1. Introduction

The Earth's atmospheric composition and surface properties are being altered by human activities. Some of the alterations are as large or larger than natural atmosphere and surface changes, even compared with natural changes that have occurred over hundreds of thousands of years. There is concern that these human-made alterations could substantially alter the Earth's climate, which has led to the United Nations Framework Convention on Climate Change [*United Nations*, 1992] with the agreed objective "to achieve stabilization of greenhouse gas concentrations in the atmosphere at a level that would prevent dangerous anthropogenic interference with the climate system."

The Earth's climate system has great thermal inertia, requiring at least several decades to adjust to a change of climate forcing [*Hansen et al.*, 1984]. Anthropogenic physical infrastructure giving rise to changes of atmospheric composition, such as power plants and transportation systems, also has a time constant for change of several decades. Thus there is a need to anticipate the nature of anthropogenic climate change and define the level of change constituting dangerous interference with nature. Simulations with global climate models on the century time scale provide a tool for addressing that need. Climate models used for simulations of future climate must be tested by means of simulations of past climate change.

We carry out climate simulations using GISS atmospheric modelE documented by *Schmidt et al.* [2006], hereafter *modelE* [2006]. Specifically, we attach the model III version of atmospheric modelE to the computationally efficient ocean model of *Russell et al.* [1995]. This coupled model and its climate sensitivity have been documented by a large set of simulations carried out to investigate the "efficacy" of various climate forcings [*Hansen et al.*, 2005a], hereafter *Efficacy* [2005]. We use the same model III here for transient climate simulations for 1880-2100, with a few simulations extended to 2300.

We made calculations for each of ten individual climate forcings for the period 1880-2003, as well as for all forcings acting together. The simulations using all ten forcings were extended into the future using scenarios of atmospheric composition defined by the Intergovernmental Panel on Climate Change [*IPCC*, 2001] and two scenarios defined by *Hansen and Sato* [2004]. The simulations for 1880-2003 with individual forcings are described by *Hansen et al.* [2006b]. Extensive diagnostics and convenient graphics for all of the runs are available at data.giss.nasa.gov/modelE/transient. Diagnostics for extended runs with all forcings acting at once are also available from the official IPCC repository (www-pcmdi.llnl.gov/ipcc/about_ipcc.php).

Section 2 defines the climate model and summarizes principal known deficiencies. Section 3 defines time-dependent climate forcings that we employ. Section 4 examines simulated global temperature change and specific regional climate change issues. Section 5 compares trends of actual climate forcings and those in the scenarios. Section 6 summarizes the relevance of our results to the basic question: is there still time to avoid dangerous human interference with climate?

## 2. Climate Model

### 2.1. Atmospheric Model

The atmospheric model employed here is the 20-layer version of GISS *modelE* [2006] with 4°x5° horizontal resolution. This resolution is coarse, but use of second-order moments for numerical differencing improves the effective resolution for the transport of tracers. The model top is at 0.1 hPa. Minimal drag is applied in the stratosphere, as needed for numerical stability, without gravity wave modeling. Stratospheric zonal winds and temperature are generally realistic [Figure 17 in *Efficacy*, 2005], but the polar lower stratosphere is as much as 5-10°C too cold in the winter and the model produces sudden stratospheric warmings at only a quarter of the observed frequency. Model capabilities and limitations are described in *Efficacy* [2005] and *modelE* [2006]. Deficiencies are summarized below (section 2.4).

### 2.2. Ocean Representation

In this paper we use the dynamic ocean model of *Russell et al.* [1995]. One merit of this ocean model is its efficiency, as it adds negligible computation time to that for the atmosphere when the ocean horizontal resolution is the same as that for the atmosphere, as is the case here. There are 13 ocean layers of geometrically increasing thickness, four of these in the top 100 m. The ocean model employs the KPP parameterization for vertical mixing [*Large et al.*, 1994] and the Gent-McWilliams parameterization for eddy-induced tracer transports [*Gent et al.*, 1995; *Griffies*, 1998]. The resulting *Russell et al.* [1995] ocean model produces a realistic thermohaline circulation [*Sun and Bleck*, 2006], but yields unrealistically weak El Nino-like variability as a result of its coarse resolution.

Interpretation of climate simulations and observed climate change is aided by simulations with the identical atmospheric model attached to alternative ocean representations. We make calculations with the same time-dependent 1880-2003 climate forcings of this paper and same atmospheric model attached to: (1) ocean A, which uses observed sea surface temperature (SST) and sea ice (SI) histories of *Rayner et al.* [2003]; (2) ocean B, the Q-flux ocean [*Hansen et al.*, 1984; *Russell et al.*, 1985], with specified horizontal ocean heat transports inferred from the ocean A control run and diffusive uptake of heat anomalies by the deep ocean; (3) ocean C, the dynamic ocean model of *Russell et al.* [1995]; (4) ocean D, the *Bleck* [2002] HYCOM ocean model. Results for ocean A and B are included in *Hansen et al.* [2006b] and on our (GISS) web site, results for ocean C are in this paper and our web site, and results for ocean D will be presented elsewhere and on our web site.

### 2.3. Model Sensitivity

The model has sensitivity 2.7°C for doubled $CO_2$ when coupled to the Q-flux ocean [*Efficacy*, 2005], but 2.9°C when coupled to the *Russell et al.* [1995] dynamical ocean. The slightly higher sensitivity with ocean C became apparent when the model run was extended to 1000 years, as the sea ice contribution to climate change became more important relative to other feedbacks as the high latitude ocean temperatures approached equilibrium. The 2.9°C sensitivity corresponds to ~0.7°C per W/m$^2$. In the coupled model with the *Russell et al.* [1995] ocean the response to a constant forcing is such that 50% of the equilibrium response is achieved in ~25 years, 75% in ~150 years, and the equilibrium response is approached only after several hundred years. Runs of 1000 years and longer are available on the GISS web site. The model's climate sensitivity of 2.7-2.9°C for doubled $CO_2$ is well within the empirical range of 3±1°C for doubled $CO_2$ that has been inferred from paleoclimate evidence [*Hansen et al.*, 1984, 1993; *Hoffert and Covey*, 1992].

### 2.4. Principal Model Deficiencies

*ModelE* [2006] compares the atmospheric model climatology with observations. Model shortcomings include ~25% regional deficiency of summer stratus cloud cover off the west coast of the continents with resulting excessive absorption of solar radiation by as much as 50 W/m$^2$, deficiency in absorbed solar radiation and net radiation over other tropical regions by typically 20 W/m$^2$, sea level pressure too high by 4-8 hPa in the winter in the Arctic and 2-4 hPa too low in all seasons in the tropics, ~20% deficiency of rainfall over the Amazon basin, ~25% deficiency in summer cloud cover in the western United States and central Asia with a corresponding ~5°C excessive summer warmth in these regions. In addition to the inaccuracies in the simulated climatology, another shortcoming of the atmospheric model for climate change studies is the absence of a gravity wave representation, as noted above, which may affect the nature of interactions between the troposphere and stratosphere. The stratospheric variability is less than observed, as shown by analysis of the present 20-layer 4°x5° atmospheric model by J. Perlwitz [personal communication]. In a 50-year control run Perlwitz finds that the interannual variability of seasonal mean temperature in the stratosphere maximizes in the region of the subpolar jet streams at realistic values, but the model produces only six sudden stratospheric warmings (SSWs) in 50 years, compared with about one every two years in the real world.

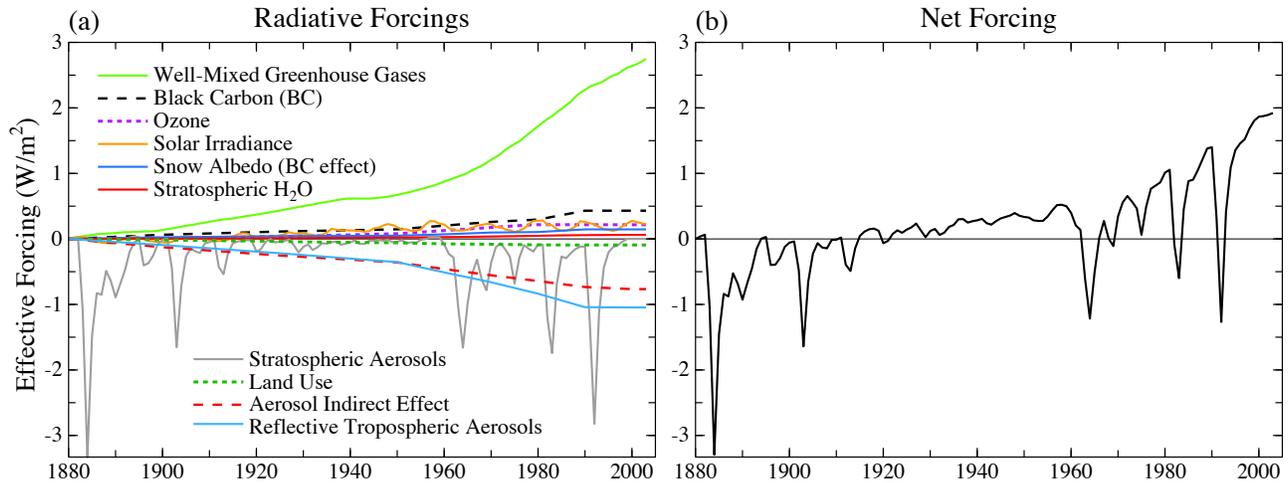

**Figure 1.** Effective global climate forcings (Fe) employed in our global climate simulations, relative to their values in 1880. Use of Fe avoids exaggerating the importance of BC and $O_3$ forcings.

The coarse resolution Russell ocean model has realistic overturning rates and inter-ocean transports [*Sun and Bleck*, 2006], but tropical SST has less east-west contrast than observed and the model yields only slight El Nino-like variability [Figure 17, *Efficacy*, 2005]. Also the Southern Ocean is too well-mixed near Antarctica [*Liu et al.*, 2003], deep water production in the North Atlantic Ocean does not go deep enough, and some deep-water formation occurs in the Sea of Okhotsk region, probably because of unrealistically small freshwater input there in the model III version of modelE. Global sea ice cover is realistic, but this is achieved with too much sea ice in the Northern Hemisphere and too little sea ice in the Southern Hemisphere, and the seasonal cycle of sea ice is too damped with too much ice remaining in the Arctic summer, which may affect the nature and distribution of sea ice climate feedbacks.

Despite these model limitations, in IPCC model inter-comparisons the model used for the simulations reported here, i.e, modelE with the Russell ocean, fares about as well as the typical global model in the verisimilitude of its climatology. Comparisons so far include the ocean's thermohaline circulation [*Sun and Bleck*, 2006], the ocean's heat uptake [*Forest et al.*, 2006], the atmosphere's annular variability and response to forcings [*Miller et al.*, 2006], and radiative forcing calculations [*Collins et al.*, 2006]. The ability of the GISS model to match climatology, compared with other models, varies from being better than average on some fields (radiation quantities, upper tropospheric temperature) to poorer than average on others (stationary wave activity, sea level pressure).

## 3. Climate Forcings

Climate forcings driving our simulated climate change during 1880-2003 arise from changing well-mixed greenhouse gases (GHGs), ozone ($O_3$), stratospheric $H_2O$ from methane ($CH_4$) oxidation, tropospheric aerosols, specifically, sulfates, nitrates, black carbon (BC) and organic carbon (OC), a parameterized indirect effect of aerosols on clouds, volcanic aerosols, solar irradiance, soot effect on snow and ice albedos, and land use changes. Global maps of each of these forcings for 1880-2000 are provided in *Efficacy* [2005].

Figure 1 shows the time dependence of the global mean effective forcings. Changes in the quantitative values of these forcings that occur with alternative forcing definitions, which are small in most cases, are discussed in *Efficacy* [2005] and in conjunction with the transient simulations for 1880-2003 [*Hansen et al.*, 2006b]. The predominant forcings are due to GHGs and aerosols, including the aerosol indirect effect. Ozone forcing is significant on the century time scale, and the more uncertain solar forcing may also be important. Volcanic effects are large on short time scales and temporal clustering of volcanoes contributes to decadal variability. Soot effect on snow and ice albedos and land use change are small global forcings, but can be large on regional scales. *Efficacy* [2005], this paper, and detailed simulations for 1880-2003 [*Hansen et al.*, 2006b] all use GHG forcings as defined in Figure 1.

GHG climate forcing, including $O_3$ and $CH_4$-derived stratospheric $H_2O$, is Fe ~ 3 W/m$^2$. Our partly subjective estimate of uncertainty, including imprecision in gas amounts and radiative transfer is ~±15%, i.e., ±0.45 W/m$^2$. Comparisons with line-by-line radiation calculations [*Collins et al.*, 2006] suggest that the $CO_2$, $CH_4$ and $N_2O$ forcings in Model III are accurate within several percent, but the CFC forcing may be 30-40% too large. If that correction is needed, it will reduce our estimated greenhouse gas forcing to Fe ~ 2.9 W/m$^2$.

Stratospheric aerosol forcing following the 1991 Mount Pinatubo volcanic eruption is probably accurate within 20%, based on a strong constraint provided by satellite measurements of planetary radiation budget [*Wong et al.*, 2004], as illustrated in Figure 11 of *Efficacy* [2005]. The uncertainty increases for earlier eruptions, reaching about ±50% for Krakatau in 1883. Between large eruptions prior to the satellite era, when small eruptions might escape detection, minimum stratospheric aerosol forcing uncertainty was ~0.5 W/m$^2$.

Tropospheric aerosols are based on emissions estimates and aerosol transport modeling, as described by *Koch* [2001]. Their indirect effect is a parameterization based on empirical effects of aerosols on cloud droplet number concentration [*Menon and Del Genio*, 2006] as described in *Efficacy* [2005]. Aerosol forcings are defined in detail by *Hansen et al.* [2006b]. The net direct aerosol forcing is Fe = -0.60 W/m$^2$ and the total aerosol forcing is Fe = -1.37 W/m$^2$ for 1880-2003. Our largely subjective

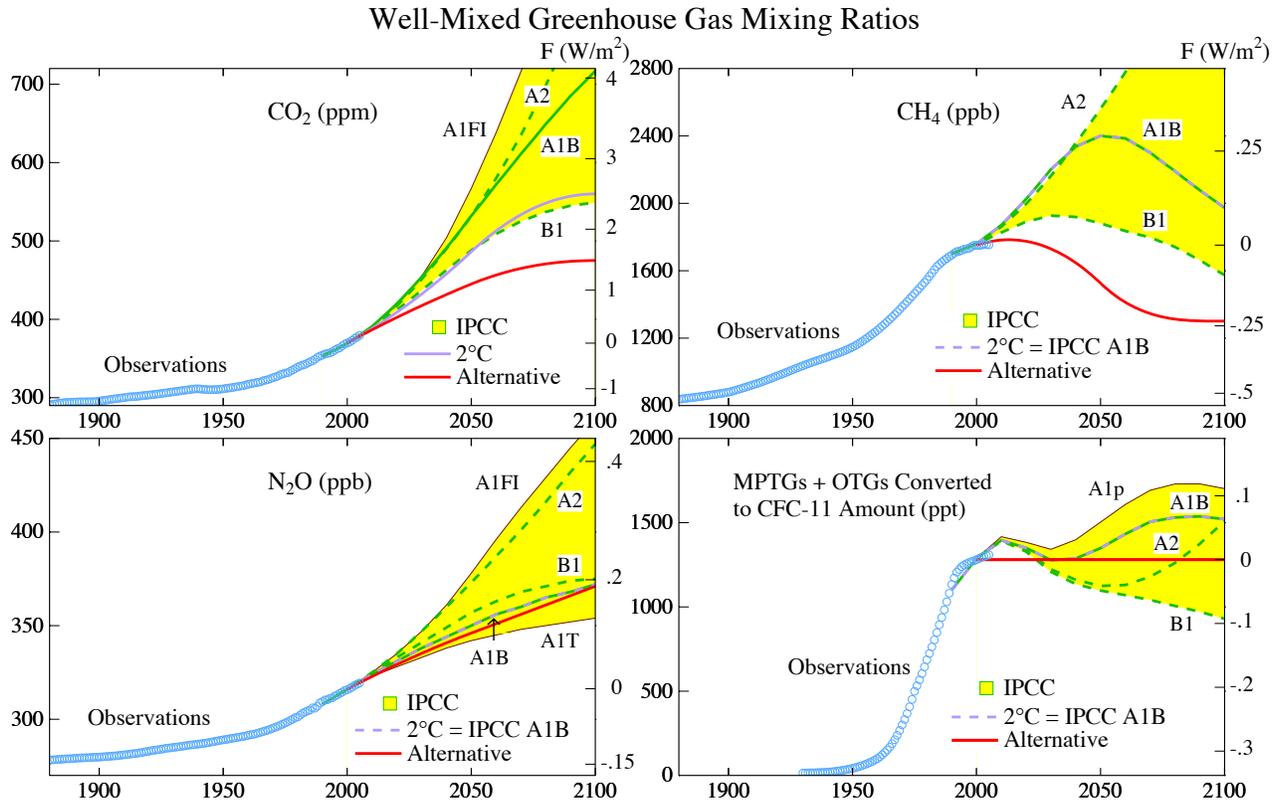

**Figure 2.** Observed greenhouse gas amounts as tabulated by *Hansen and Sato* [2004] and scenarios for the 21st century. Colored area delineates extreme *IPCC* [2001] scenarios. Alternative and 2°C scenarios are from Table 2 of *Hansen and Sato* [2004]. MPTGs and OTGs are Montreal Protocol trace gases and other trace gases [*Hansen and Sato*, 2004]. Forcings on right hand scales are adjusted forcings, Fa, relative to values in 2000.

estimate of the uncertainty in the net aerosol forcing is at least 50%.

The sum of all forcings is Fe ~ 1.90 W/m$^2$ for 1880-2003. However, the net forcing is evaluated more accurately from an ensemble of simulations carried out with all forcings present at the same time [*Efficacy*, 2005], thus accounting for any non-linearity in the combination of forcings and minimizing the effect of noise (unforced variability) in the climate model runs. All forcings acting together yield Fe ~ 1.75 W/m$^2$.

Uncertainty in the net forcing for 1880-2003 is dominated by the aerosol forcing uncertainty, which is at least 50%. Given our estimates, we must conclude that the net forcing is uncertain by ~ 1 W/m$^2$. Therefore the smallest and largest forcings within the range of uncertainty differ by more than a factor of three, primarily because of the absence of accurate measurements of aerosol direct and indirect forcings.

One implication of the uncertainty in the net 1880-2003 climate forcing is that it is fruitless to try to obtain an accurate empirical climate sensitivity from observed global temperature change of the past century. However, paleoclimate evidence of climate change between periods with well-known boundary conditions (forcings) provides a reasonably precise measure of climate sensitivity: 3±1°C for doubled $CO_2$ [*Hansen et al.*, 1984, 1993; *Hoffert and Covey*, 1992]. Thus we conclude that our model sensitivity of 2.9°C for doubled $CO_2$ is reasonable.

Figure 1b indicates that, except for occasional large volcanic eruptions, the GHG climate forcing has become the dominant global climate forcing during the past few decades. GHG dominance is a result of the slowing growth of anthropogenic aerosols, as increased aerosol amounts in developing countries have been at least partially balanced by decreases in developed countries. This dominance of GHG forcing, with a net forcing that may be approaching the equivalence of a 1% increase in solar irradiance, implies that global temperature change should now be rising above the level of natural climate variability.

Furthermore, we can anticipate that the dominance of GHG forcing over aerosol and other forcings will be all the more true in the future [*Andreae et al.*, 2005]. There is little expectation that developing countries will allow aerosol amounts to continue to grow rapidly, as their aerosol pollution is already hazardous to human health and technologies to reduce emissions are available. If future GHG amounts are anywhere near the projections in typical IPCC scenarios, GHG climate forcings will be dominant in the future. Thus, we suggest, it is possible to make meaningful projections of future climate change despite uncertainties in past aerosol forcings.

*IPCC* [2001] defines a broad range of scenarios for future greenhouse gas amounts. This range is shown nominally by the colored area in Figure 2, which is bordered by the two IPCC scenarios that give the largest and smallest greenhouse gas amounts in 2100. We carry out climate simulations for three IPCC scenarios: A2, A1B and B1. A2 and B1 are, respectively, near the maximum and minimum of the range of *IPCC* [2001] scenarios, and A1B is known as the IPCC midrange baseline scenario.

We also carry out simulations for the "alternative" and "2°C" scenarios (Figure 2) defined in Table 2 of *Hansen and*

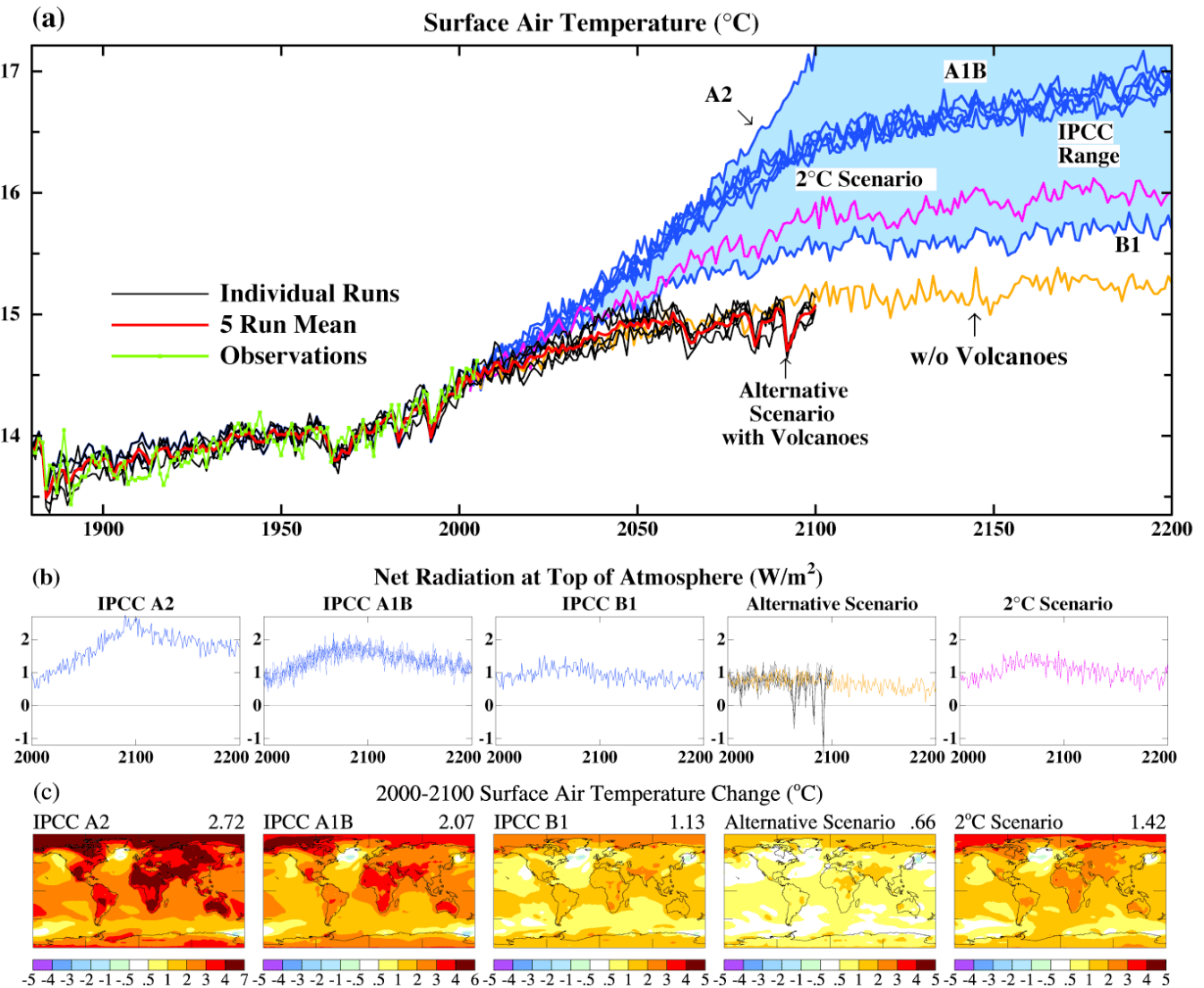

**Figure 3.** Global mean surface air temperature for several scenarios calculated by our coupled climate model as extensions of the 1880-2003 simulations for "all forcings". Forcings beyond 2003 for these scenarios are defined in Figure 2. Stratospheric aerosols in 2010-2100 are the same as in 1910-2000 for the case including future volcanoes. Tropospheric aerosols are unchanging in the 21st century.

*Sato* [2004]. $CO_2$ increases 75 ppm during 2000-2050 in the alternative scenario; $CH_4$ decreases moderately, enough to balance a steady increase of $N_2O$; chlorofluorocarbons 11 and 12 decrease enough to balance the increase of all other trace gases, an assumption that is handled by having all these gases remain constant after 2000. $CO_2$ peaks at 475 ppm in 2100 in the alternative scenario, while $CH_4$ decreases to 1300 ppb. The alternative scenario is designed to keep the added forcing at about 1 W/m² in 2000-2050 and another 0.5 W/m² in 2050-2100. With a nominal climate sensitivity of ¾°C per W/m² and slowly declining greenhouse gas amounts after 2100, this scenario prevents global warming from exceeding 1°C above the global temperature in 2000, a level that *Hansen* [2004, 2005a,b] has argued would constitute "dangerous anthropogenic interference" with global climate.

*O'Niell and Oppenheimer* [2002] suggest 2°C added global warming as a limit defining "dangerous anthropogenic interference". Thus we also made climate simulations for a scenario expected to approach but not exceed that limit. The "2°C" scenario has $CO_2$ peak at 560 ppm in 2100 and other greenhouse gases follow the IPCC midrange scenario A1B (see www.giss.nasa.gov/data/simodel/ghgases/).

### 4. Climate Simulations

Simulations for the historical period, 1880-2003, warrant careful comparison with observations and analysis of effects of individual forcings. Results are described in detail by *Hansen et al.* [2006b] and model diagnostics are available at data.giss.nasa.gov/modelE/transient. We briefly summarize results relevant to analysis of the level of future global warming that might constitute dangerous climate interference.

The model driven by the forcings of Figure 1 simulates observed 1880-2003 global temperature change reasonably well, as crudely apparent in Figure 3a. An equally good fit to observations probably could be obtained from a model with larger sensitivity (than 2.9°C for doubled $CO_2$) and smaller net forcing, or a model with smaller sensitivity and larger forcing, but paleoclimate evidence constrains climate sensitivity, as mentioned above. Also our model responds realistically to the known short-term forcing by Pinatubo volcanic aerosols [*Hansen et al.*, 2006b], and simulated global warming for the past few decades, when increasing greenhouse gases were the dominant forcing (Figure 1), is realistic [*Hansen et al.*, 2005b, 2006b].

**Table 1.** Forcings and Surface Air Temperature Response to Different Forcings for Several Periods in the Future

| Forcing Agent | Run Name | Fi | Fa | $\Delta T_s$ [2000 to Year Below] | | | |
|---|---|---|---|---|---|---|---|
| | | 2000-2100 | | 2050 | 2100 | 2200 | 2300 |
| *IPCC Scenarios* | | | | | | | |
| A2 | E3IP_A2 | 6.67 | 6.23 | 1.06 | 2.71 | 3.33 | 3.70 |
| A1B | E3IP_A1B | 4.72 | 4.36 | 1.04 | 1.99 | 2.51 | 2.75 |
| B1 | E3IP_B1 | 2.61 | 2.38 | 0.73 | 1.20 | 1.36 | 1.54 |
| *GISS Scenarios* | | | | | | | |
| Alternative S. w/o Volcanoes | E3ALT | 1.59 | 1.44 | 0.50 | 0.80 | 0.89 | — |
| Alternative S. with Volcanoes | E3ALTV | — | — | 0.53 | 0.56 | — | — |
| 2°C Scenario w/o Volcanoes | E3_2Cx | 3.08 | 2.85 | 0.81 | 1.46 | 1.66 | — |
| *Constant Scenario* | | | | | | | |
| 2003 Forcings | E3Af8xa | 0.05 | 0.05 | 0.20 | 0.28 | — | — |

The largest discrepancies in simulated 1880-2003 surface temperature change are deficient warming in Eurasia and excessive warming of the tropical Pacific Ocean. *Hansen et al.* [2006b] present evidence that the deficient Eurasian warming may be due to an excessive anthropogenic aerosol optical depth in that region. The lack of notable observed warming in the tropical Pacific could be due to increased frequency or intensity of La Ninas [*Cane et al.*, 1997], a characteristic that our present model would not be able to capture regardless of whether it was a forced or unforced change.

Overall, simulated climate change does not agree in all details with observations, but such agreement is not expected given unforced climate variability, uncertainty in climate forcings, and current model limitations. However, the climate model does a good job of simulating global temperature change from short time-scale (volcanic aerosol) to century time-scale forcings. This provides incentive to examine model results for evidence of dangerous human-made climate effects and to investigate how these effects depend upon alternative climate forcing scenarios.

**4.1. Global Temperature Change**

We carry out climate simulations for the 21st century and beyond for *IPCC* [2001] scenarios A2, A1B, and B1 and the "alternative" and "2°C" scenarios of *Hansen et al.* [2000] and *Hansen and Sato* [2004]. Simulations are continued beyond 2100 with forcings fixed at 2100 levels as requested by IPCC, although the alternative and 2°C scenarios per se have a slow decline of human-made forcings after 2100.

Figure 3a and Table 1 show the simulated global mean surface air temperature change. The A2 and B1 simulations are single runs, while A1B is a 5-member ensemble. A 5-member ensemble of runs is carried out for the alternative scenario including future volcanoes, the 2010-2100 volcanic aerosols being identical to those of 1910-2000 in the updated *Sato et al.* [1993] index. Single runs are made for the alternative and 2°C scenarios without future volcanoes.

Global warming is 0.80°C between 2000 and 2100 in the alternative scenario (for the 25-year running mean at 2100; it is 0.66°C based on linear fit) and 0.89°C between 2000 and 2200 (0.77°C based on linear fit) with atmospheric composition fixed after 2100. The Earth is out of energy balance by ~ ¾ W/m² in 2100 (~0.4 W/m² in 2200) in this scenario (Figure 3b), so global warming eventually would exceed 1°C if atmospheric composition remained fixed indefinitely. The alternative scenario of *Hansen and Sato* [2004] has GHGs decrease slowly after 2100, thus keeping warming < 1°C. Global warming in all other scenarios already exceeds 1°C by 2100, ranging from 1.2°C in scenario B1 to 2.7°C in scenario A2.

Note the slow decline of the planetary energy imbalance after 2100 (Figure 3b), which reflects the shape of the surface temperature response to a climate forcing. Figure 4d in *Efficacy* [2005] shows that 50% of the equilibrium response is achieved within 25 years, but only 75% after 150 years, and the final 25% requires several centuries. This behavior of the coupled model occurs because the deep ocean continues to take up heat for centuries. Verification of this behavior in the real world requires data on deep ocean temperature change. In the model, heat storage associated with this long tail of the response curve occurs mainly in the Southern Ocean. Measured ocean heat storage in the past decade [*Willis et al.*, 2004; *Lyman et al.*, 2006] presents limited evidence of this phenomenon, but the record is too short and the measurements too shallow for confirmation. Ongoing simulations with modelE coupled to the current version of the *Bleck* [2002] ocean model do not show such deep mixing of heat anomalies, at least not so rapidly.

Figure 3c compares the global distribution of 21st century temperature change in the five scenarios. Warming in the alternative scenario is < 1°C in most regions, reaching 1°C only in the Arctic and a few continental regions. The weakest IPCC forcing, B1, has almost twice the warming of the alternative scenario, with more than 2°C in the Arctic. The strongest IPCC forcing, A2, yields four times the global warming of the alternative scenario, with annual warming exceeding 4°C on large areas of land and in the Arctic. We compare these warmings to observed change and interannual variability in the next section.

**4.2. Regional Climate Change**

Figure 4 compares climate change simulated for the present century with climate change and climate variability that humans and the environment experienced in the past century. Figure 4a is the observed change (based on linear trend) of seasonal (winter and summer) surface temperature over the past century and the local standard deviation, σ, of seasonal mean temperature about its 100-year mean. The range within which seasonal-mean temperature has varied in the past millennium is probably no more than twice the range shown in Figure 4a for the past cen-tury [*Mann et al.*, 2003].

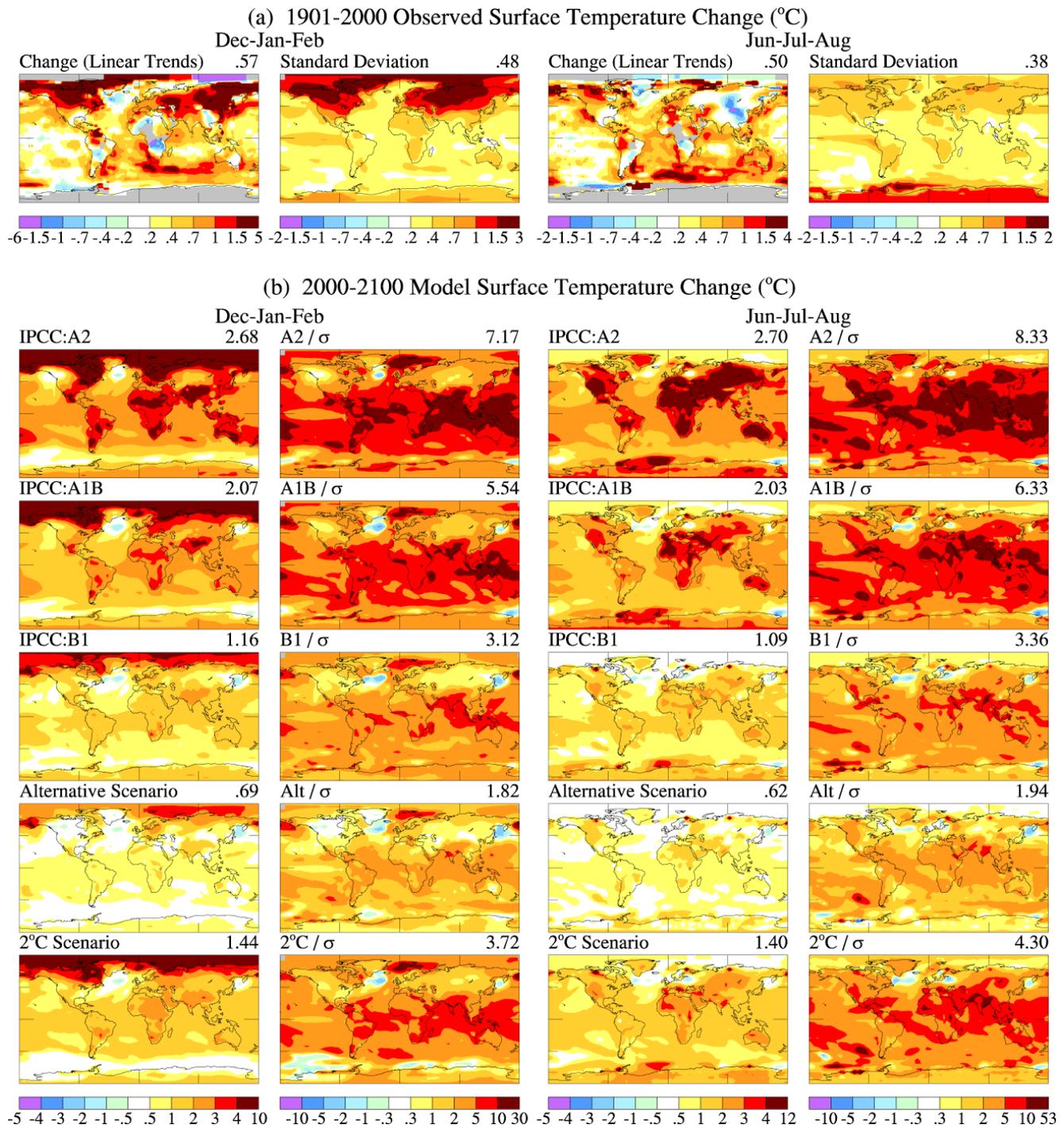

**Figure 4.** (a) Observed seasonal (DJF and JJA) surface temperature change in the past century based on local linear trends and the standard deviation about the local 100-year mean temperature, (b) simulated 21st century seasonal temperature change for five GHG scenarios and the ratio of the simulated change to the observed 20th century standard deviation.

Figure 4b shows the simulated change of seasonal mean temperature this century and the ratio of this change to observed local temperature variability. Warming in the alternative scenario is typically $2\sigma$ or less, i.e., the average seasonal mean temperature at the end of the 21st century will be at a level that is occasionally experienced in today's climate. Warming in the 2°C and IPCC B1 scenarios is typically $4\sigma$. Warming in the IPCC A2 and A1B scenarios, commonly called "business-as-usual" (BAU) scenarios, is typically $5\text{-}10\sigma$.

Ecosystems, wildlife, and humans would be subjected in the BAU scenarios to conditions far outside their local range of experience. We suggest that $5\text{-}10\sigma$ changes of seasonal temperature are *prima facie* evidence that the BAU scenarios extend well into the range of "dangerous anthropogenic interference". Figure 4 provides a general perspective on the magnitude of regional climate change expected for a given climate forcing scenario. Additional perspective on the practical significance of regional climate change is obtained by considering three specific cases: the Arctic, tropical storms originating in the Tropical Atlantic Ocean, and the ocean in the vicinity of ice shelves.

**4.2.1. Arctic climate change.** Recent warming in the Arctic is having notable effects on regional ecology, wildlife, and indigenous peoples [*ACIA*, 2004]. Unforced climate variability is especially large in the Arctic (Figure 4, top row), where modeled variability is similar in magnitude to observed variability (Figure 11 in *Hansen et al.*, 2006b). Observed variability includes the effect of forcings, but, at least in the model, forcings are not yet large enough to have noticeable effect on the regional standard deviation about the long-term trend. Prior to recent warming, the largest fluctuation of Arctic temperature was the brief strong Arctic warming around 1940, which *Johannessen et al.* [2004] and *Delworth and Knutson* [2000] argue was an unforced fluctuation associated with the Arctic Oscillation and a resulting positive anomaly of ocean heat inflow into that region. Forcings such as solar variability [*Lean and Rind*, 1998] or volcanoes [*Overpeck et al.*, 1997] may contribute to Arctic variability, but we do not expect details of observed climate change in the first half of the 20th century to be matched by model simulations because of the large unforced variability. However, the magnitude of simulated Arctic warming over the entire century [illustrated by *Hansen et al.*, 2006b] is consistent with observations. Given the attribution of at least a large part of global warming to human-made forcings [*IPCC*, 2001], high latitude amplification of warming in climate models and paleoclimate studies, and the practical impacts of observed climate change [*ACIA*, 2004], Arctic climate change warrants special attention.

Early energy balance climate models revealed a "small ice cap instability" at the pole [*Budyko*, 1969; *North*, 1984], which implied that, once sea ice retreated to a critical latitude, all remaining ice would be lost rapidly without additional forcing. This instability disappears in climate models with a seasonal cycle of radiation and realistic dynamical energy transports, but a vestige remains: the snow/ice albedo feedback makes sea ice cover in summer and fall sensitive to moderate increase of climate forcings. The Arctic was ice-free in the warm season during the Middle Pliocene when global temperature was only 2-3°C warmer than today [*Crowley*, 1996; *Dowsett et al.*, 1996].

Satellite data indicate a rapid decline, ~9%/decade, in perennial Arctic sea ice since 1978 [*Comiso*, 2002], raising the question of whether the Arctic has reached a 'tipping point' leading inevitably to loss of all warm season sea ice [*Lindsay and Zhang*, 2005]. Indeed, some experts suggest that "…there seem to be few, if any, processes or feedbacks that are capable of altering the trajectory toward this 'super interglacial' state" free of summer sea ice [*Overpeck et al.*, 2005].

Could the Greenland ice sheet survive if the Arctic were ice-free in summer and fall? It has been argued that not only is ice sheet survival unlikely, but its disintegration would be a wet process that can proceed rapidly [*Hansen*, 2004, 2005a,b]. Thus an ice-free Arctic Ocean, because it may hasten melting of Greenland, may have implications for global sea level, as well as the regional environment, making Arctic climate change centrally relevant to definition of dangerous human interference.

Are there realistic scenarios that might avoid large Arctic warming and an ice-free Arctic Ocean? *Efficacy* [2005] and *Shindell et al.* [2006] suggest that non-$CO_2$ forcings are the cause of a substantial portion of Arctic climate change, and thus reduction of these pollutants would make stabilization of Arctic climate more feasible. Investigation of this topic should include simulations of future climate for plausible changes of forcings that affect the Arctic. Although that is beyond the scope of our present paper, we can make research suggestions on the basis of simulations that we have done for 1880-2003.

The top row of Figure 5 shows observed 1880-2003 surface temperature change and our simulated temperature change for 'all forcings', in both cases the Arctic warming being about twice the global warming. The second row shows the results of simulations including only the $CO_2$ forcing (left) and only $CH_4$ plus tropospheric $O_3$ (right). $CO_2$ and $CH_4 + O_3$ yield comparable global and Arctic warmings. The sum of the responses to $CO_2$ and $CH_4 + O_3$ exceeds either observed warming or the simulated warming for 'all forcings', because the real world and 'all forcings' include negative forcings, which are due primarily to aerosols.

The lower left of Figure 5 shows the simulated temperature change due to the effect of black carbon on snow albedo. In this simulation, described in detail by *Hansen et al.* [2006b], a moderate snow albedo forcing (~0.05 W/m²) is assumed. The BC producing this snow albedo effect also has direct and indirect aerosol forcings, and the sources producing BC also produce OC. The lower right of Figure 5 shows the simulated temperature change due to the combination of $CH_4 + O_3 + BC + OC$, including aerosol direct and indirect forcings and the snow albedo effect. The global warming is about the same as that due to $CH_4 + O_3$, as the warming by BC and its snow albedo effect is not much larger than the cooling by OC and the indirect effect of BC and OC. Nevertheless, simulated Arctic warming is more than 1°C with all of these forcings present.

If $CO_2$ growth in the 21st century is kept as small as in the alternative scenario, additional Arctic warming is only ~1°C (Figure 3c). Thus, in that case, reduction of some of the pollutants considered in Figure 5 may make it possible to keep further Arctic warming under 1°C and thus probably avoid loss of all sea ice. On the other hand, if $CO_2$ growth follows a BAU scenario, the impact of reducing the non-$CO_2$ forcings will be small by comparison and probably inconsequential.

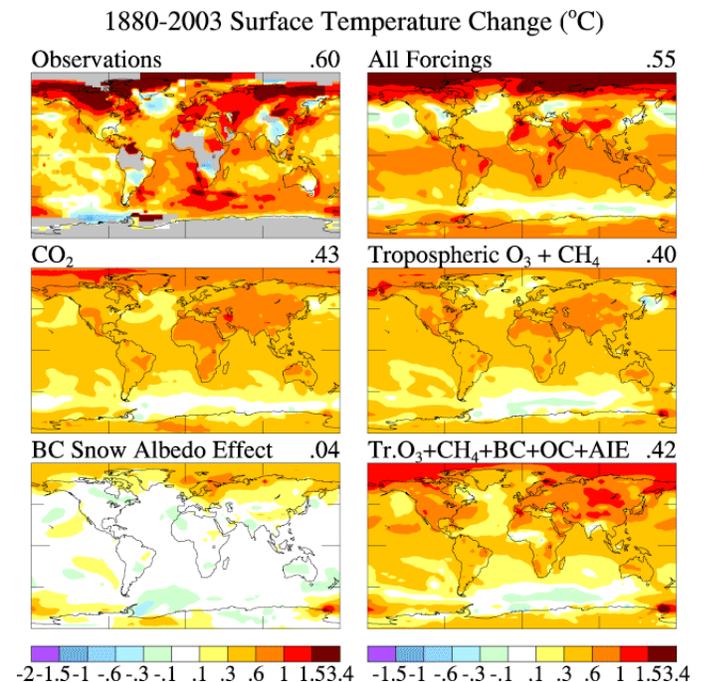

**Figure 5.** Surface temperature change based on local linear trends for observations and for simulations that employ different combinations of transient 1880-2003 forcings. $CH_4$ forcing includes its indirect effect on stratospheric $H_2O$. Snow albedo effect has 1880-2003 Fa ~ 0.05 W/m². Results in lower right include direct effects of black carbon (BC) and organic carbon (OC) aerosols from fossil fuels and biomass, their aerosol indirect effects (AIE), and the snow albedo effect.

We suggest that the conclusion that a 'tipping point' has been passed, such that it is not possible to avoid a warm-season ice-free Arctic, with all that might entail for regional climate and the Greenland ice sheet, is not warranted yet. Better information is needed on the present magnitude of all anthropogenic forcings and on the potential for their reduction. If $CO_2$ growth is kept close to that of the alternative scenario, and if strong efforts are made to reduce positive non-$CO_2$ forcings, it may be possible to minimize further Arctic climate change.

**4.2.2. Tropical climate change.** The tropical Atlantic Ocean is the spawning grounds for tropical storms and hurricanes that strike the United States and Caribbean nations. Many local and large-scale factors affect storm activity and there are substantial correlations of storm occurrence and severity with external conditions such as the Southern Oscillation and Quasi-Biennial Oscillation [*Gray*, 1984; *Emanuel*, 1987; *Henderson-Sellers et al.*, 1998; *Goldenberg, et al.*, 2001; *Trenberth*, 2005]. Two factors especially important for Atlantic hurricane activity are SST at 10-20°N, the main development region (MDR) for Atlantic tropical storms [*Goldenberg, et al.*, 2001], and the absence of strong vertical wind shear that would inhibit cyclone formation. In addition, SST and ocean temperature to a few hundred meters depth along the hurricane track play a role in storm intensity, because warm waters enhance the potential for the moist convection that fuels the storm, while cooler waters at depth stirred up by the storm can dampen its intensity.

Some measures of Atlantic hurricane intensity and duration increased in recent years, raising concern that global warming may be a factor in this trend [*Bell et al.*, 2005; *Webster et al.*, 2005, *Emanuel*, 2005]. However, *Gray* [2005] and the director of the United States National Hurricane Center [*Mayfield*, 2005], while acknowledging a connection between ocean temperature and hurricanes, reject the suggestion that global warming has contributed to the recent storm upsurge, citing instead natural cycles of Atlantic Ocean temperature. Indeed, multi-decadal variations of Atlantic Ocean temperatures are found in instrumental data [*Kushnir*, 1994], in paleoclimate proxy temperatures [*Mann et al.*, 1998], and in coupled atmosphere-ocean climate simulations without external forcing [*Delworth and Mann*, 2000]. However, *Mann and Emanuel* [2006] argue that SST fluctuations in the MDR associated with the Atlantic Multi-decadal Oscillation are statistical artifacts, and that practically all SST variability in the region can be attributed to competing trends in greenhouse gases and aerosols.

Climate forcings also contribute to ocean temperature change, so it is of interest to compare modeled temperature change due to forcings with observed temperature change in regions and season relevant to tropical storms. *Bell et al.* [2005] define an Accumulated Cyclone Energy (ACE) index that accounts for the combined strength and duration of tropical storms of hurricanes originating in the Atlantic Ocean. This index [Fig. 4.5 of *Bell et al.*, 2005] was generally low during 1970-1994. For the past decade the ACE index has been a factor 2.4 higher than in 1970-1994. Perhaps not coincidentally, the index had peaks near 1980 and 1990, when global temperature also had peaks, and the index was low during the few years affected by the 1991 Pinatubo global cooling. The past decade, when ACE is highest, is the warmest time in the past century and a period with rapid uptake of heat by the ocean [*Levitus et al.*, 2005; *Willis et al.*, 2004; *Hansen et al.*, 2005b; *Lyman et al.*, 2006].

Figure 6a shows SST anomalies in 1995-2005, the time of high hurricane intensity, relative to 1970-1994, when Atlantic Ocean and Gulf of Mexico hurricanes were weaker. Observations have warming ~0.2°C in the Gulf of Mexico and ~0.45°C in the MDR. The ensemble mean of the simulation with standard "all forcings" has warming about 0.35°C in both regions, suggesting that much, perhaps most, of observed warming in these critical regions is due to the forcings that drive the climate model. Increasing GHGs, the principal forcing driving the model toward warming (Figure 1), are known accurately, but other forcings have large uncertainties. Thus we performed additional simulations in which the most uncertain forcings were altered to test the effect of uncertainties and to provide more fodder for examining statistical significance.

The lower three rows in Figure 6a (and 6c) show results for three 5-member ensembles of model runs with altered tropospheric aerosol and solar forcings, as described in detail by *Hansen et al.* [2006b]. In AltAer1 anthropogenic sulfate aerosols have the identical time dependence as in the standard experiment (Figure 1) but the anthropogenic increase is reduced by 50%. AltAer2 adds to AltAer1 a doubling of the temporal increase of biomass burning aerosols. AltSol replaces the solar irradiance history of *Lean* [2000], which includes both long-term and solar cycle variations, with the time series of *Lean et al.* [2002], which retains only the solar change due to the 11-year Schwabe cycle. These alternative forcings all yield ensemble-mean SST warmings in the MDR comparable to that with standard forcings.

Figure 6b shows the 1995-2005 temperature anomalies in the five individual runs with "all forcings", each of which yields warming in both the Gulf of Mexico and the MDR. Figure 6c shows the standard deviations σ* among successive 10-year periods relative to the preceding 25 years within the available relevant period of record (1900-1994). As expected, the model has too little variability at low latitudes, but it is more realistic at middle and high latitudes (the small 'observed' SST variability at the highest latitudes is an artificial result of specifying a fixed SST if any sea ice is present). σ* is ~0.15°C in the Gulf of Mexico in observations and model. Observed σ* is ~0.25°C in the MDR, but only about half that large in the model.

In summary, the warming in the model in recent decades is due to the assumed forcings, and *Hansen et al.* [2006b] present evidence that the magnitude of the model's response to forcings is realistic on time scales from that for individual volcanic eruptions to multidecadal GHG increases. The period 1970-2005 under discussion with regard to hurricanes is the time when forcings are known most accurately, and during that period anthropogenic GHGs were the dominant forcing. Although unforced fluctuations undoubtedly contribute to Atlantic Ocean temperature change, the expected GHG warming is comparable in magnitude to observed warming and must be at least a significant contributor to that warming.

We conclude that the definitive assertion of *Gray* [2005] and *Mayfield* [2005], that human-made GHGs play no role in the Atlantic Ocean temperature changes that they assume to drive hurricane intensification, is untenable. Specifically, the assertions that (1) hurricane intensification of the past decade is due to changes in SST in the Atlantic Ocean, and (2) global warming cannot have had a significant role in the hurricane intensification of the past decade, are mutually inconsistent. On the contrary, although natural cycles play a role in changing Atlantic SST, our model results indicate that, to the degree that hurricane intensification of the past decade is a product of increasing SST in the Atlantic Ocean and the Gulf of Mexico, human-made GHGs probably are a substantial contributor, as also concluded by Mann and Emanuel [2006]. *Santer et al.* [2006] have obtained similar conclusions by examining the results of 22 climate models.

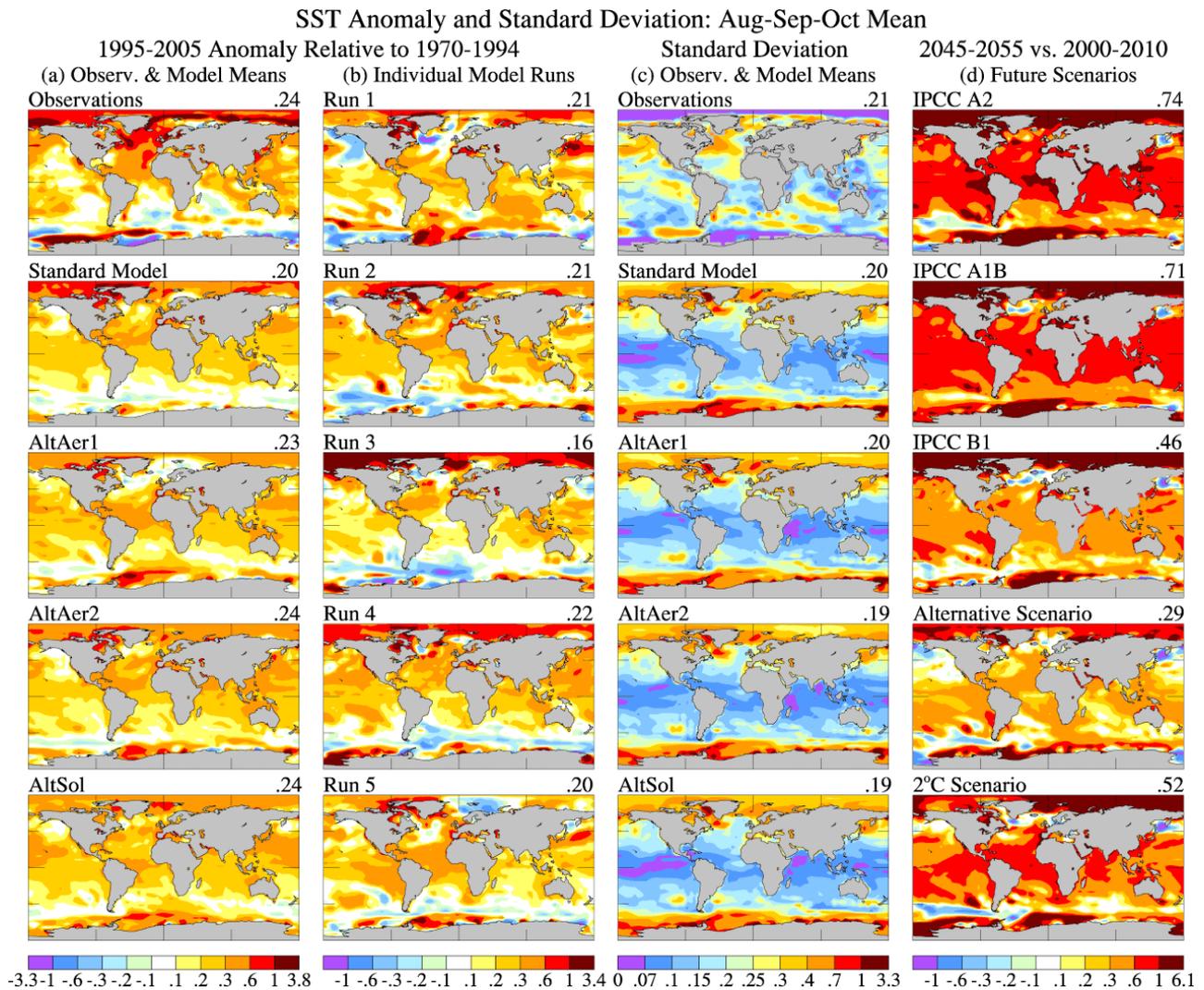

**Figure 6.** (a) 1995-2005 Aug-Oct SST anomalies relative to 1970-1994 base period, the top map for observations and next four for several 'all forcing' ensembles, (b) 1995-2005 Aug-Oct SST anomalies relative to 1970-1994 for individual ensemble members of the standard 'all forcings' simulation, (c) standard deviations among successive decadal means relative to immediately preceding 25 years, (d) simulated 2045-2055 Aug-Oct SST anomalies relative to the present (2001-2009 mean).

Figure 6d shows the additional SST warming during hurricane season by mid-century in the three IPCC, alternative, and 2°C scenarios, relative to the present. SST is only one of the environmental factors that affect hurricanes, but there is theoretical [*Emanuel*, 1987] and empirical evidence [*Emanuel*, 2005; *Webster et al.*, 2005] that higher SST contributes to increased maximum strength of tropical cyclones. The salient point that we note in Figure 6d is that, despite the fact that warming "in the pipeline" in 2004 is the same for all scenarios, the alternative scenario already at mid-century has notably less warming than the other scenarios. This result refutes the common statement that constraining climate forcings has a negligible effect on expected warming this century. Divergence of warming among the scenarios is even greater in later decades, as the growth of climate forcing declines rapidly in the alternative scenario.

**4.2.3. Ice sheet and methane hydrate stabilities.** Perhaps the greatest threat of catastrophic climate impacts for humans is the possibility that warming may cause one or more of the ice sheets to become unstable, initiating a process of disintegration that is out of humanity's control. We focus on ice sheet stability, because of the potential for sea level change this century. However, ice sheets are related to the stability of methane hydrates in permafrost and in marine sediments, because ice sheet disintegration and methane release are both affected by surface warming at high latitudes and by ocean warming at the depths that affect ice shelves and shallow methane hydrate deposits. Also, a substantial positive climate feedback from methane release could make it difficult to achieve the relatively benign alternative scenario, thus affecting the ice sheet stability issue as well as other climate impacts. Most methane hydrates in marine sedi-HANSEN ET AL.: DANGEROUS HUMAN-MADE INTERFERENCE WITH CLIMATEments are at depths beneath the ocean floor that would require centuries for destabilization [*Kvenvolden and Lorenson*, 2001; *Archer*, 2006], but the anthropogenic thermal burst in ocean temperatures is likely to persist for centuries. Furthermore, the methane in shallow sediments, i.e., at or near the sea floor, might begin to yield a positive climate feedback this century if substantial warm thermal anomalies penetrate to the ocean bottom.

The Greenland and Antarctic ice sheets contain enough water to raise sea level about 70 m, with Greenland and West Antarctica each containing about 10% of the total and East Antarctica holding the remaining 80%. *IPCC* [2001] central estimates for sea level change in the 21st century include negligible contributions from Greenland and Antarctica, as they presume that melting of ice sheet fringes will be largely balanced by ice sheet growth in interior regions where snowfall increases.

However, Earth's history shows that the long-term response to global warming includes substantial melting of ice and sea level rise. During the penultimate interglacial period, ~130,000 years ago, when global temperature may have been as much as ~1°C warmer than in the present interglacial period, sea level was 4 ± 2 m higher than today [*McCulloch and Esat*, 2000; *Thompson and Goldstein*, 2005], demonstrating that today's sea level is not particularly favored. However, changes of sea level and global temperature among recent interglacial periods are not large compared to the uncertainties, and ice sheet stability is affected not only by global temperature but also by the geographical and season distribution of solar irradiance, which differ from one interglacial period to another. Therefore, it is difficult to use the last interglacial period as a measure of the sensitivity of sea level to global temperature. The most recent time with global temperature ~3°C greater than today (during the Pliocene, ~ 3 million years ago) had sea level 25 ± 10 m higher than today [*Barrett et al.*, 1992; *Dowsett et al.* 1994; *Dwyer et al.*, 1995], suggesting that, given enough time, a BAU level of global warming could yield huge sea level change. A principal issue is thus the response time of ice sheets to global warming.

It is commonly assumed, perhaps because of the time scale of the ice ages, that the response time of ice sheets is millennia. Indeed, ice sheet models designed to interpret slow paleoclimate changes respond lethargically, i.e., on millennial time scales. However, perceived paleoclimate change may reflect more the time scale for changes of forcing, rather than an inherent ice sheet response time. Ice sheet models used for paleoclimate studies generally do not incorporate all the physics that may be critical for the wet process of ice sheet disintegration, e.g., modeling of the ice streams that channel flow of continental ice to the ocean, including effects of melt percolation through crevasses and moulins, removal of ice shelves by the warming ocean, and dynamical propagation inland of the thinning and retreat of coastal ice. *Hansen* [2005a,b] argues that, for a forcing as large as that in BAU climate forcing scenarios, a large ice sheet response would likely occur within centuries, and ice melt this century could yield sea level rise of 1 m or more and a dynamically changing ice sheet that is out of our control.

Two regional climate changes particularly relevant to ice sheet response time are: (1) surface melt on the ice sheets, and (2) melt of submarine ice shelves that buttress glaciers draining the ice sheets. *Zwally et al.* [2002] showed that increased surface melt on Greenland penetrated ice sheet crevasses and reached the ice sheet base, where it lubricated the ice-bedrock interface and accelerated ice discharge to the ocean. *Van den Broeke* [2005] found that unusually large surface melt on an Antarctic ice shelf preceded and probably contributed significantly to collapse of the ice shelf. Thinning and retreat of ice shelves due to warming ocean water works in concert with surface melt. *Hughes* [1972, 1981] and *Mercer* [1978] suggested that floating and grounded ice shelves extending into the ocean serve to buttress outlet glaciers, and thus a warming ocean that melts ice shelves could lead to rapid ice sheet shrinkage or collapse. Confirmation of the effectiveness of these processes has been obtained in the past decade from observations of increased surface melt, ice shelf thinning and retreat, and acceleration of glaciers on the Antarctic Peninsula [*De Angelis and Skvarca*, 2003; *Rignot et al.*, 2004; *van den Broeke*, 2005], West Antarctica [*Rignot*, 1998; *Shepherd et al.*, 2002, 2004; *Thomas et al.*, 2004], and Greenland [*Abdalati et al.*, 2001; *Zwally et al.*, 2002; *Thomas et al.*, 2003], in some cases with documented effects extending far inland. Thus we examine here both projected surface warming on the ice sheets and warming of nearby ocean waters.

Summer surface air temperature change on Greenland in the 21st century is only 0.5-1°C with alternative scenario climate forcings (Figure 4), but it is 2-4°C for the IPCC BAU scenarios (A2 and A1B). Simulated 21st century summer warmings in West Antarctica (Figure 4) differ by a similar factor, being ~0.5°C in the alternative scenario and ~2°C in the BAU scenarios.

Submarine ice shelves around Antarctica and Greenland extend to ocean depths of 1 km and deeper [*Rignot and Jacobs*, 2002]. Figure 7b shows simulated zonal mean ocean temperature change versus depth for the past century and the 21st century. Despite near absence of surface warming in the Antarctic Circumpolar Current, warming of a few tenths of a degree is found at depths from a few hundred to 1500 m, consistent with observations in the past half-century [*Gille*, 2002; *Aoki et al.*, 2003]. Modeled warming is due to GHGs, with less warming when all forcings are included.

Maps of simulated temperature change for the top 1360 m of the ocean (Figure 7a) show warming around the entire circumference of Greenland, despite cooling in the North Atlantic. This warming at ocean depth differs by about a factor of two between the alternative and BAU scenarios, less than the factor ~4 difference in summer surface warming over Greenland and Antarctica, consistent with the longer response time of the ocean. Although we cannot easily convert temperature increase into rate of melting and sea level rise, it is apparent that the BAU scenarios pose much greater risk of large sea level rise than the alternative scenario, as discussed in Section 6.

Surface and submarine temperature change are also relevant to the stability of methane hydrates in permafrost and in ocean sediments. Recent warming is already beginning to cause release of methane from thawing permafrost in Siberia [*Walter et al.*, 2006; *Zimov et al.*, 2006], but this methane release is not yet large enough to yield an overall increase of atmospheric $CH_4$, as shown in section 5. Projected 21st century summer warming in the Northern Hemisphere permafrost regions is typically a factor of five larger in BAU scenarios than in the alternative scenario (Figure 4), suggesting that the threat of a significant positive climate feedback is much greater in the BAU scenarios.

Methane hydrates in ocean sediments are an even larger potential source of methane emissions [*Archer*, 2006]. It is possible that much of the temperature rise in extreme global warming events, such as the approximately 6°C that occurred at the Paleocene-Eocene boundary about 55 million years ago [*Bowen et al.*, 2006], resulted from catastrophic release of methane from methane hydrates in ocean sediments. It may require

11 of 21

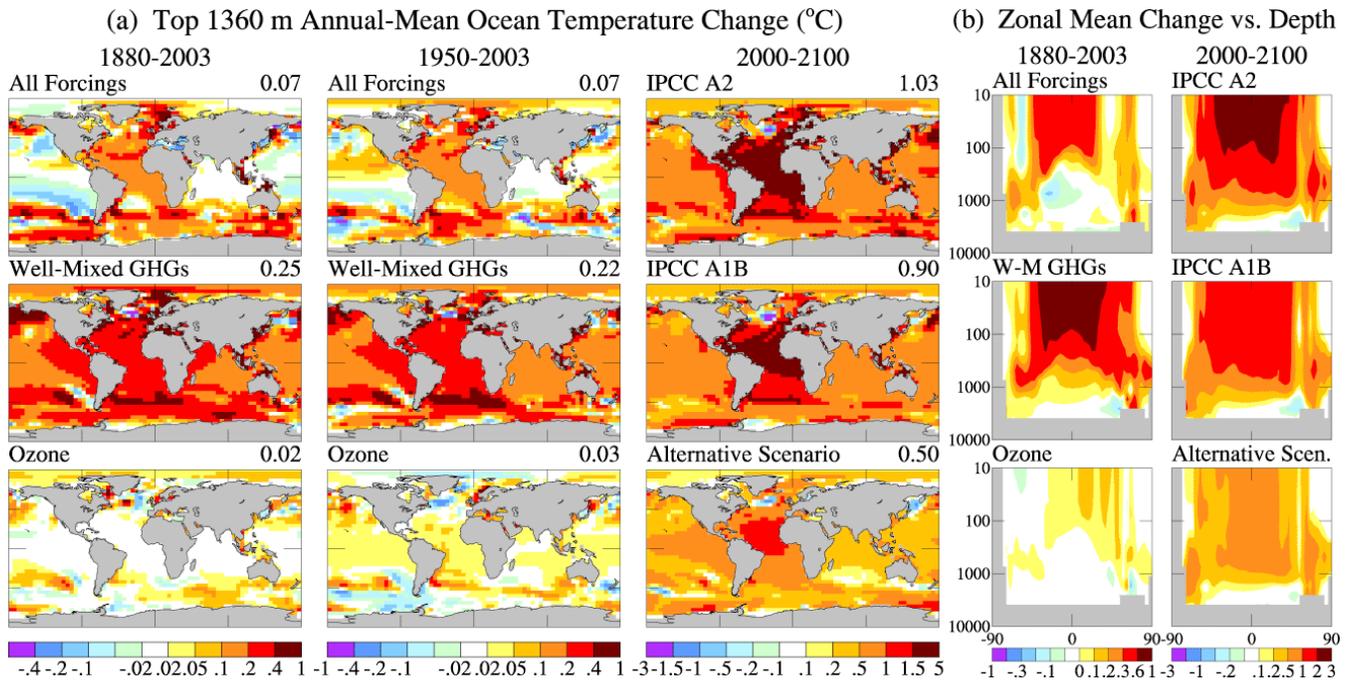

**Figure 7.** (a) Temperature change in the top 10 layers (1360 m) of the ocean for three periods and three forcings, (b) zonal-mean ocean temperature change for these forcings. Positive and negative temperature intervals are symmetric about zero.

many centuries for ocean warming to substantially impact these methane hydrates [*Archer*, 2006]. Calculations of global warming enhancement from methane hydrates based on diffusive or upwelling-diffusion ocean models suggest that the methane hydrate feedback is small on the century time scale [*Harvey and Huang*, 1995]. However, the positive thermal anomalies that we simulate with our dynamic ocean model at great ocean depths at high latitudes (Figure 7) are much larger near the ocean bottom than we would obtain with diffusive mixing of anomalies (as in our Q-flux model), suggesting that further attention to the possibility of methane hydrate release is warranted. With our present lack of understanding, we can perhaps only say with reasonable confidence that the chance of significant methane hydrate feedback is greater with BAU scenarios than with the alternative scenario, and that empirical evidence from prior interglacial periods suggests that large methane hydrate release is unlikely if global warming is kept within the range of recent interglacial periods.

### 5. Actual GHG Trends versus Scenarios

Global and regional climate changes simulated for IPCC scenarios in the 21st century are large in comparison with observed climate variations in the 20th century. This raises the question: is it plausible for global climate forcing to follow a path with a smaller forcing than those in the IPCC scenarios?

### 5.1. Ambient GHG Amounts

Figure 8a compares GHG scenarios and observations, the latter being an update of *Hansen and Sato* [2004] (see www.giss.nasa.gov/data/simodel/ghgases), who define the data sources and methods of obtaining global means from station measurements. Figure 8b shows the corresponding annual growth rates. Figure 8c and 8d compare observed climate forcings and their growth rates with the IPCC, 2°C, and alternative scenarios.

Observed $CO_2$ falls close to all scenarios, which do not differ much in early years of the 21st century. $CO_2$ has large year-to-year variations in its growth rate due to variations in terrestrial and ocean sinks, as well as biomass burning and fossil fuel sources. However, we are able to draw conclusions in section 5.3 about the realism of $CO_2$ scenarios by comparing emission scenarios with real world data on fossil fuel $CO_2$ emission trends, for which there are reasonably accurate data.

Overall, growth rates of the well-mixed non-$CO_2$ forcings fall below IPCC scenarios (Figure 8a,b). Growth of $CH_4$ falls below any IPCC scenario and even below the alternative scenario. Observed $N_2O$ falls slightly below all scenarios. The sum of MPTGs (Montreal Protocol Trace Gases) and OTGs (Other Trace Gases) falls between the IPCC scenarios and the alternative scenario (which was defined at a later time than the IPCC scenarios, when more observational data were available). The estimated forcing by MPTGs is based on measurements of 10 of these gases, as delineated by *Hansen and Sato* [2004], all of which are growing more slowly than in the IPCC scenarios, with the few significant unmeasured MPTGs assumed to follow the lowest IPCC scenario. OTGs with reported data include HFC-134a and $SF_6$, for which measurements are close to or slightly below IPCC scenarios.

Figures 8c and 8d show that the net forcing by well-mixed GHGs for the past few years has been following a course close to that of the alternative scenario. However, it should not be assumed that forcings will remain close to the alternative scenario, because, as shown below, fossil fuel $CO_2$ emissions now substantially exceed those in the alternative scenario.

The alternative scenario was defined [*Hansen et al.*, 2000] as a potential goal under the assumption of concerted global efforts to simultaneously (1) reduce air pollution (for human health, climate and other reasons), and (2) stabilize $CO_2$ emissions initially and begin to achieve significant emission reductions before mid-century, such that added $CO_2$ forcing is held to ~1 W/m² in 50 years and ~1.5 W/m² in 100 years. Thus the alternative scenario assumes that $CH_4$ amount will peak within a decade and then decline enough to balance continued growth of $N_2O$. It also assumes that the decline of $CH_4$ and other $O_3$ precursors will decrease tropospheric $O_3$ enough to balance the small increase in forcing (several hundredths of a W/m$^2$) expected due to recovery of halogen-induced stratospheric $O_3$ depletion.

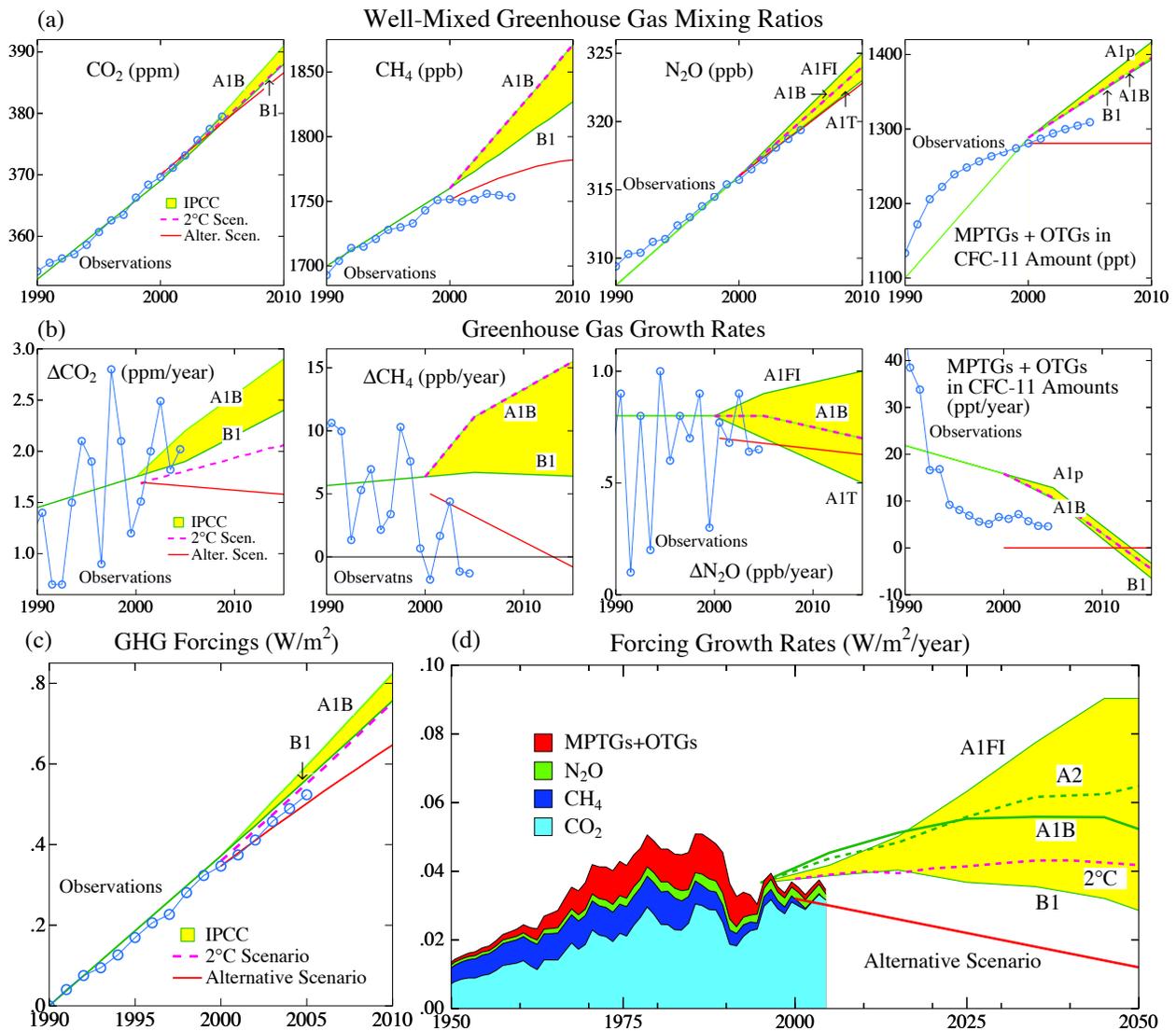

**Figure 8.** (a) Greenhouse gas amounts, (b) growth rates, and (c) resulting forcings and (d) forcing growth rates for *IPCC* [2001], "alternative", and "2°C" scenarios [*Hansen et al.*, 2000; *Hansen and Sato*, 2004]. Observations are 5-year running means of data described by *Hansen and Sato* [2004] and in the text. Data for recent decades are those reported by the NOAA Earth System Research Laboratory, Global Monitoring Division (www.cmdl.noaa.gov). MPTGs and OTGs without available observations (4.9% of the MPTG + OTG estimated forcing in 2000) were assumed to follow IPCC scenario A1B, which probably exaggerates their forcing, as all measured MPTGs and OTGs are currently falling below their IPCC scenarios.

The most demanding requirement of the alternative scenario is that added $CO_2$ forcing be held to ~1 W/m$^2$ in 50 years and ~1.5 W/m$^2$ in 100 years. The scenario in our present climate simulations, for example, has annual mean $CO_2$ growth of 1.7 ppm/year at the end of the 20$^{th}$ century declining to 1.3 ppm/year by 2050. The plausibility of the alternative scenario thus depends critically upon fossil fuel $CO_2$ emissions.

### 5.2. Fossil Fuel $CO_2$ Emissions

Global fossil fuel $CO_2$ *emissions* change by only a few percent per year and are known with sufficient accuracy that we can draw conclusions related to atmospheric $CO_2$ trends, despite large year-to-year variability in *atmospheric* $CO_2$ growth. In recent decades the increase of atmospheric $CO_2$, averaged over several years, has been ~58% of fossil fuel emissions [*IPCC*, 2001; *Hansen*, 2005b]. Carbon cycle models suggest that this 'airborne fraction' of $CO_2$ is unlikely to decrease this century if emissions increase, indeed, most models yield an increasing airborne fraction as terrestrial $CO_2$ uptake is likely to decrease [*Cox et al.*, 2000; *Friedlingstein, et al.*, 2003; Fung et al., 2005; *Matthews et al.*, 2005; *Jones et al.*, 2005].

We use the historical emissions data set of *Marland et al.* [2006] supplemented with data for 2004-2005 from *British Petroleum* [2006]. The data are imperfect, e.g., the temporary reduction of coal use in China in the late 1990s may have been exaggerated by reporting problems [*Sinton*, 2001], but effort has been made to make the data as temporally consistent as possible [*Marland et al.*, 2006]. We include estimates for $CO_2$ from gas flaring (0.3% of $CO_2$ emissions) and cement production (3.8% of emissions in 2005). We normalize *British Petroleum* [2006] data for each fossil fuel such that it coincides with the final year of *Marland et al.* [2006] data (2003 in present analysis).

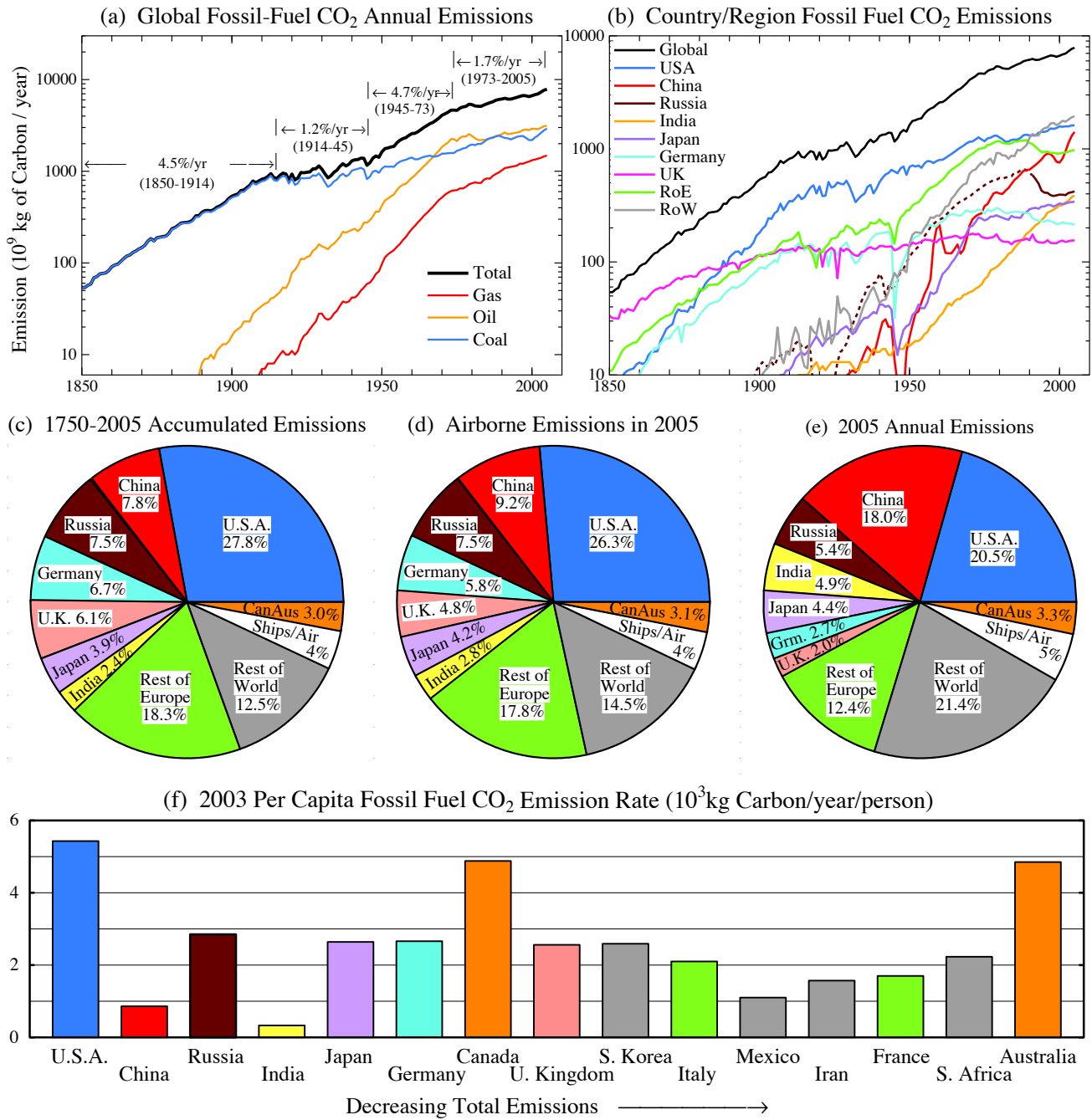

**Figure 9.** Fossil fuel annual emissions in terms of (a) fuel type, and (b) source country or region. (c) is accumulated 1750-2005 emissions by source region, (d) is the contribution to anthropogenic $CO_2$ remaining in the air in 2005 based on an analytic fit to the Bern carbon cycle model, (e) is 2005 emissions, and (f) is per capita emissions of the 15 largest sources in 2003. Data sources: Carbon Dioxide Information Analysis Center, Oak Ridge National Laboratory for 1751-2003 and *British Petroleum* [2006] for 2004-2005. Data for Russia are 60% of USSR for 1850-1991 and Russian Federation thereafter.

Global fossil fuel $CO_2$ emissions grew at >4%/year for most of the century preceding 1973, but since 1973 at ~1.7%/year and for the past decade at ~2%/year (Figure 9a). Coal has produced 50% of fossil fuel $CO_2$ emissions through 2005 (oil 37%, gas 13%), and, with the recent uptick in its use (Figure 9a), coal seems poised to retake from oil the role of the largest current $CO_2$ source, calling into question the inevitability of continued decarbonization of global energy [*Ausubel*, 1996, 2000], at least in the near-term.

Continued growth of $CO_2$ emissions at 1.7%/year doubles emissions in ~40 years and is close to IPCC BAU growth. IPCC scenarios B1, A1B, A2 and A1F1 require 50-year constant growth rates of 1.1%, 1.7%, 1.8% and 2.4%, respectively, to reach their 2050 emission rates.

Figure 9b reveals near congruence of the global growth curve for $CO_2$ emissions with that for the United States, perhaps reflecting the fact that the United States has long been a technology leader and suggesting that it may be difficult for global emissions to flatten out and decline unless the United States follows such a course. However, Figure 9b also shows the rapid rise of emissions from China, India and 'Rest of World' (principal components being Southeast Asia, Middle East, Canada, Australia, South America, Africa), whose combined portion of global emissions is now 48% and growing. $CO_2$ emissions from Europe and the Russian Federation have declined since 1990. Japan had rapid emissions growth from 1950 until the oil embargo and price rise of 1973, but nearly flat subsequent emissions as a result of concerted efficiency efforts and growing use of nuclear power [*Okabe*, 2002], illustrating the compatibility of strong economic growth with constrained emissions.

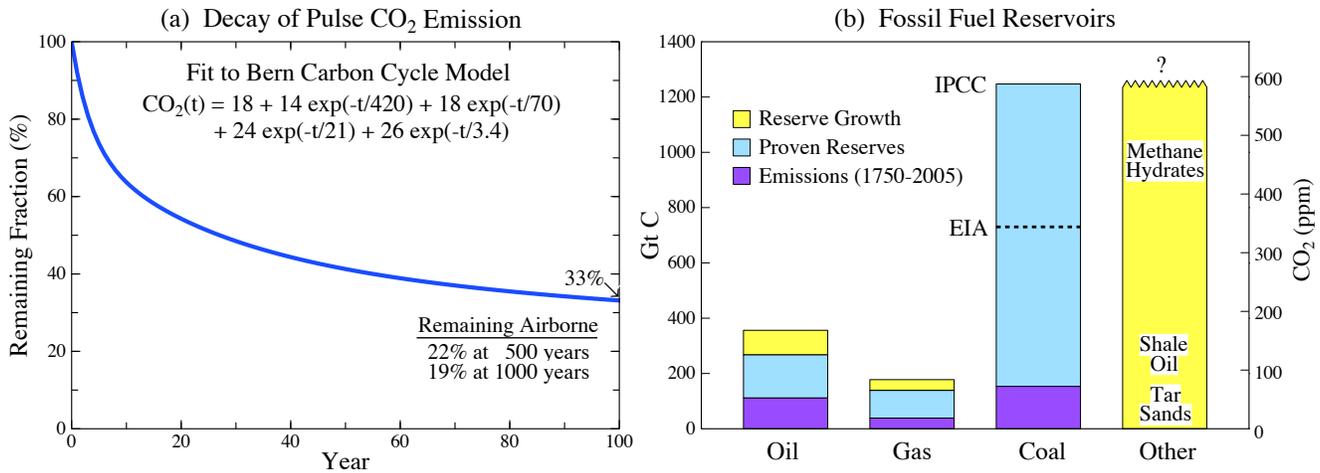

**Figure 10.** (a) Decay of a small pulse of $CO_2$ added to today's atmosphere, based on analytic approximation to the Bern carbon cycle model [*Joos et al.*, 1996; *Shine et al.*, 2005], (b) fossil fuel emissions to date based on sources in Figure 9, and proven reserves and estimated economically recoverable reserve growth based on *EIA* [2005] and, in the case of coal, *IPCC* [2001].

The rationale for the alternative scenario is that a leveling of $CO_2$ emissions is practical in the near term based on existing technologies, especially via improved energy efficiencies, and that reduced long-term emissions could be achieved via $CO_2$ sequestration, renewable energies, improved nuclear power, and/or new non-fossil energy sources. *Pacala and Socolow* [2004] provide a quantitative discussion of several technologies that together provide sufficient 'wedges' to reduce near-term $CO_2$ emissions growth to 0%. However, leveling off of emissions will not occur without concerted efforts to adopt required technologies. In view of the growing proportion of emissions from developing countries, it will also require international cooperation and sharing of technologies.

Failure of current emission trends to meet the alternative scenario can be quantified using the empirical fact that the $CO_2$ 'airborne fraction' (ratio of observed atmospheric $CO_2$ increase to fossil fuel $CO_2$ emissions), averaged over several years, has remained fixed at nearly 60% for half a century. 60% of current fossil fuel emissions yields an annual airborne $CO_2$ increment of 2 ppm, which should thus be viewed as the present underlying growth rate of atmospheric $CO_2$, even though actual increases fluctuate strongly from year to year. The alternative scenario permits $CO_2$ growth of 75 ppm between 2000 and 2050 (85 ppm for IPCC estimates of forcing per ppm) to yield a 50-year forc-ing of 1 W/m², and it assumes that this might be achieved, for example, via $CO_2$ annual growth decreasing from the late 20[th] century value of 1.7 ppm/year to 1.3 ppm/year by mid-century.

Overshoot of the 1 W/m² target for added $CO_2$ forcing this half-century can be balanced, in principle, by reducing non-$CO_2$ forcings. However, the maximum likely decrease of non-$CO_2$ forcings is several tenths of 1 W/m², so continued 2%/year growth of $CO_2$ emissions for even one more decade may make the alternative scenario impractical to achieve. The action most threatening to the alternative scenario would be extensive construction of new coal-fired power plants without provision for $CO_2$ sequestration, because of the long life of power plant infrastructure. Climatic consequences of coal-fired power plants can be averted via coal gasification with $CO_2$ sequestration [*Hawkins*, 2005], but during the next 10 years that technology is only being tested. Thus the alternative scenario's near-term requirement of stabilizing $CO_2$ emissions requires emphasis on greater use of renewable energy sources and, especially, energy efficiency, which together have the potential to satisfy increased energy needs for at least 1-2 decades [*Romm et al.*, 1998; *Pacala and Socolow*, 2004; *NCEP*, 2004].

Carbon cycle facts sketched in Figure 10 aid discussion of the role of fossil fuel $CO_2$ emissions in determining future climate forcing, and specifically help define constraints on $CO_2$ emissions that are needed if the alternative scenario is to be achieved. Figure 10 (a) shows the decay of a small pulse $CO_2$ emission based on a simple analytic fit to the Bern carbon cycle model [*Joos et al.*, 1996]

$$CO_2 (\%) = 18 + 14\, e^{-t/420} + 18\, e^{-t/70} + 24\, e^{-t/21} + 26\, e^{-t/3.4}. \quad (1)$$

In (1), $t$ is time, and we slightly rounded coefficients calculated by Joos [*Shine et al.* 2005].

In this approximation of the carbon cycle, about one-third of anthropogenic $CO_2$ emissions remain in the atmosphere after 100 years (Figure 10a) and one-fifth after 1000 years. Processes not accounted for by the model, including burial of carbon in ocean sediments from perturbation of the $CaCO_3$ cycle and silicate weathering, remove $CO_2$ on longer time scales [*Archer*, 2005], but for practical purposes 500-1000 years is "forever", as it is long enough for ice sheets to respond and it is many human generations. Indeed, the function (1) should be viewed as an approximate lower bound for the portion of fossil fuel $CO_2$ emissions that remain airborne. The uptake capacity of the ocean decreases as the amount of dissolved carbon increases (the buffer factor increases) and there are uncertain but potentially large climate feedbacks that may add $CO_2$ to the air, e.g., carbon emissions from forest dieback [*Cox et al.*, 2000], melting permafrost [*Walter et al.* 2006; *Zimov et al.*, 2006], and warming ocean bottom [*Archer*, 2006].

Integration over the period 1750-2005 of the product of (1) and fossil fuel emissions of Fig. 9a yields a present airborne fossil fuel $CO_2$ amount of ~80 ppm [*Kharecha and Hansen*, 2006]. The observed atmospheric $CO_2$ increase is ~100 ppm, the difference presumably due to the net consequence of deforestation and biospheric uptake not incorporated in the carbon cycle model, and, in part, imprecision of the carbon cycle model. This calculation provides a check on the reasonableness of the carbon cycle model approximation (1), which should continue to provide useful estimates for the moderate fossil fuel emissions inherent in the alternative scenario. As mentioned above, for the greater $CO_2$ emissions of BAU scenarios equation (1) may begin to substantially underestimate airborne $CO_2$, as it excludes the nonlinearity of the ocean carbon cycle and anticipated climate feedbacks on atmospheric $CO_2$ and $CH_4$, the latter being eventually oxidized to $CO_2$.

Figure 10b suggests the key role that coal and unconventional fossil fuels (labeled "other") will play in determining future $CO_2$ levels and the attainment or non-attainment of the alternative scenario. If growth of oil and gas reserves estimated by the Energy Information Administration [*EIA*, 2005] is realistic, full exploitation of oil and gas will take airborne atmospheric $CO_2$ to ~450 ppm, assuming that coal use is phased out over the next several decades except for uses where the $CO_2$ produced can be captured and sequestered [*Kharecha and Hansen*, 2006]. Achievement of this $CO_2$ limit also requires that the massive amounts of unconventional fossil fuels (labeled 'other' in Figure 10) are exploited only if the resulting $CO_2$ is captured and sequestered.

## 6. Summary: Can We Avoid Dangerous Climate Change?

### 6.1. Global Warming: How Much is Dangerous?

*IPCC* [2001] provides an invaluable service by assessing knowledge of climate change and addressing the goal of the United Nations Framework Convention on Climate Change [Article 2, *United Nations*, 1992], which is to stabilize atmospheric composition at a level avoiding 'dangerous anthropogenic interference' with climate. *IPCC* [2001] discusses dangerous interference with the aid of a 'burning embers' diagram, in which several 'reasons for concern' are assigned colors that become increasingly red as global temperature increases, the color assignments being based on expert opinions. The diagram is sometimes taken as suggesting that global warming of 2-3°C, relative to recent temperatures, is likely to be dangerous [*IPCC*, 2001; *Smith and Hitz*, 2003; *Yohe et al.*, 2004]. *Schneider and Mastrandrea* [2005] use the burning embers for a probabilistic assessment of dangerous, with 2.85°C as the 50$^{th}$ percentile threshold.

*Hansen* [2004, 2005a,b] asserts that the dangerous level of global warming is closer to 1°C, his principal rationale being evidence from the Earth's history that greater warmings are likely to cause large sea level change, which, he argues, will occur on a time scale of centuries. We discuss the sea level issue in section 6.1.1 and regional climate effects in 6.1.2, describing additional evidence in support of the contention that a level of global warming as low as 1°C can have dangerous effects.

**6.1.1. Global temperature and sea level.** Simulated 21$^{st}$ century climate change can be used to drive ice sheet models that calculate the expected contribution of ice sheets to sea level change. *Wild et al.* [2003] include ice sheet calculations within their high-resolution (T-106) global climate model with a BAU scenario that has $CO_2$ reaching 715 ppm in 2100, finding that both the Greenland and Antarctic ice sheets grow, contributing a 12 cm/century fall to sea level change. *Huybrechts et al.* [2004] use a 3-D thermomechanical ice sheet/ice shelf model driven by climate change produced by general circulation models (ECHAM4 and HadAM3H), finding mass loss from Greenland, larger mass gain by Antarctica, with little net sea level change by 2100.

We question whether such ice sheet models adequately represent all physics that is important during the wet process of ice sheet disintegration in a warming climate. Explicit modeling of ice streams that channel flow of ice to the ocean may be needed, including effects of rainfall and melt percolation through crevasses and moulins, as well as the removal of ice shelves by the warming ocean and dynamical propagation inland of effects due to thinning and retreat of coastal ice. Increased snowfall in ice sheet interiors provides one negative feedback, but with BAU warming of several degrees Celsius over the ice sheets most of Greenland and West Antarctica would be bathed in summer melt. As the area of melt increases, ice shelves are lost, and coastal ice thins, positive feedbacks due to decreasing albedo and lowering ice surface may feed a rapid nonlinear dynamical response.

Given the difficulty of modeling ice sheet response to warming, it is useful to examine the Earth's history for evidence of ice sheet response to prior warmings. Global temperature should be a meaningful parameter for such empirical study. It is shown in *Efficacy* [2005] that there is a congruence in the spatial distribution of climate response to most global forcing mechanisms normalized by the effective (or fixed SST) forcing, with exceptions for highly localized forcings such as land-use change or biomass burning aerosols, and with limitation to comparisons for a fixed distribution of continents. This approximate equivalence of effective forcings increases the likelihood of a meaningful relationship between global temperature and climate impacts such as sea level change.

The highest sea level in the past half million years was at most ~ +5 m relative to today. A case has been made for higher sea level ~400 Kybp [*Hearty et al.*, 1999], but there are dating uncertainties and it is likely that this higher stand occurred at a much earlier time [*Droxler et al.*, 2002]. It is necessary to go back to the Pliocene, about 3 million years ago, to find sea levels clearly much greater than today. Sea level then is estimated to have been 25 ± 10 m higher than today [*Barrett et al.*, 1992; *Dowsett et al.*, 1994; *Dwyer et al.*, 1995]. Maximum global temperatures during interglacial periods of the past half million years were probably about +1°C relative to today; Antarctic temperatures in some interglacial periods reached about +2°C [*Petit et al.*, 1999], but tropical maxima were < +1°C [*Medina-Elizade and Lea*, 2005; *Hansen et al.*, 2006a]. The middle Pliocene was about +3°C relative to today [*Crowley*, 1996; *Dowsett et al.*, 1996].

Long-term global ice volume changes tend to lag global temperature change by a few thousand years [*Mudelsee*, 2001], but these changes are in response to weak forcings varying on millennial time scales [*Hays et al.*, 1976]. Nevertheless, at some ice age terminations ice volume change is rapid, e.g., the largest sea level rise following the last ice age, Meltwater Pulse 1A, when sea level rose about 20 m in 400 years, was practically synchronous with the Bolling warming [*Kienast et al.*, 2003]. Other studies provide evidence of large (~ 10 m or more) changes of sea level on "sub-orbital" time scales [*Siddall et al.*, 2003; *Potter et al.*, 2004; *Thompson and Goldstein*, 2005], i.e., sea level changes occurring much more rapidly than changes of the Earth's orbital elements.

GHG climate forcings in the IPCC BAU scenarios, such as A1FI, A2 and A1B, are far outside the range that has existed on Earth in millions of years. The rate of change of this sustained forcing exceeds that of known forcings in at least millions of years. Global warming in the BAU scenarios, with canonical climate sensitivity ~ 3°C for doubled $CO_2$, is ~ 3°C by 2100 and still rising rapidly. By 2100 the melt area on Greenland and West Antarctica and the melt rate of ice shelves, if any ice shelves remain, would be much greater than today. Thus it is unlikely that the response time for significant ice sheet change could exceed centuries, because such response times (sea level change of meters per century) have occurred during the Pleistocene with much smaller forcings.

If equilibrium sea level rise is many meters, a response time of centuries provides little consolation to coastal dwellers. They would be faced with intermittent floods associated with storms and continually rebuilding above a transient sea level. Thus we suggest that sea level change may define a low level for global warming that constitutes dangerous change, due to the large concentration of people and infrastructure along global coastlines.

Present understanding of ice sheet response to global warming does not allow sharp definition of a 'dangerous' level, but BAU scenarios are surely well into the dangerous regime. Even global warming of 1°C conceivably could produce a long-term sea level rise of several meters [*Otto-Bleisner et al.*, 2006; *Overpeck et al.*, 2006]. However, climate forcing on the ice sheets is far smaller with global warming < 1°C than with global warming 2-3°C, and the resulting slower changes of the ice sheets would allow a better chance to develop climate mitigation strategies or adapt to sea level change.

**6.1.2. Regional climate effects.** Regional climate change also yields a clear distinction between BAU scenarios with global warming ~ 3°C and an alternative scenario that keeps global warming <~ 1°C. In the specific alternative scenario we have defined, global warming is 0.80°C in the 21st century for a model with climate sensitivity 2.9°C for doubled $CO_2$. The resulting change of seasonal mean temperature this century is typically 1-2 σ, where σ is the standard deviation of seasonal temperature in the 20th century. In IPCC BAU scenarios, the average change of seasonal mean temperature is 5-10 σ.

Ecosystems, wildlife, and humans thus would be subjected in the BAU scenarios to conditions far outside their local range of experience. Perhaps humans, aided by modern technology, may adapt readily to such regional climate change. Wildlife and ecosystems, however, will not have that advantage, and their abilities to migrate may be limited by anthropogenic and geographic constraints on their locations. We suggest that 5-10σ changes of seasonal temperature are *prima facie* evidence that the BAU scenarios extend well into the range of "dangerous anthropogenic interference".

Arctic climate change provides another possible criterion for DAI. The strong positive feedback between sea ice area and surface albedo makes the Arctic one of the most sensitive regions on Earth to global warming. In the middle Pliocene, with global temperature 2-3°C warmer than today, the Arctic was ice-free in the warm season. Such drastic climate change would have deleterious effects on wildlife and indigenous people [*ACIA*, 2004], constituting what many people would agree is dangerous anthropogenic interference with nature.

Thus, from the perspective of the Arctic, IPCC BAU scenarios again yield dangerous climate change. Positive feedbacks, however, can be minimized by keeping the net forcing small. We suggest that, if additional $CO_2$ growth is kept small, as in the alternative scenario, and if some of the climate forcings that are most effective in the Arctic are reversed, i.e., reduced in absolute magnitude, it may be possible to limit or even eliminate further Arctic warming. This would require large reductions of $CH_4$, tropospheric $O_3$, and other pollutants, but such reductions would have ancillary benefits [*West et al.*, 2006; *Air Pollution Workshop*, www.giss.nasa.gov/meetings/pollution2005].

**6.1.3. Climate feedbacks.** The climate sensitivity of ¾°C per W/m² (about 3°C for doubled $CO_2$) is inferred for the case in which GHGs and ice sheet area are treated as forcings [*Hansen*, 2005b]. If climate change is small enough, say in the range of the interglacial periods of the past million years, the areas of the Greenland and Antarctica ice sheets should not change significantly, and the positive feedback of increasing GHGs with increasing global temperature is small [+18 ppm $CO_2$ per °C, +49 ppb $CH_4$ per °C, +15 ppb $N_2O$ per °C, *Hansen and Sato*, 2004]. However, with BAU scenarios the expected global warming extends far outside the range of recent interglacial periods, raising the possibility of much larger feedbacks such as destabilization of methane hydrates (section 4.2.3).

Although we lack sufficient information for quantitative analysis, the existence of climate feedbacks appears to present the possibility of a dichotomy between climate futures such as the alternative scenario, which keeps global warming less than 1°C, and BAU scenarios. If the warming is less than 1°C, it appears that strong positive feedbacks are not unleashed, judging from recent Earth history. On the other hand, if global warming gets well out of this range, there is a possibility that positive feedbacks could set in motion climate changes far outside the range of recent experience. The ability of the ocean to absorb human-made $CO_2$ decreases as the emissions increase [*Archer*, 2005], there is a possibility that the terrestrial biosphere could even become a source of $CO_2$ [*Cox et al.* 2000], and even a potential that large amounts of methane could be released from methane hydrates, as discussed in section 4.2.3.

**6.2. Can We Avoid Dangerous Climate Change?**

Global warming in "business-as-usual" (BAU) climate forcing scenarios, for climate sensitivity consistent with paleoclimate data, is at least 2-3°C by 2100 (relative to 2000) and still increasing rapidly. Implications include an ice-free Arctic in the warm season, other regional climate changes outside the range of historical experience, and initiation of ice sheet changes that presage future sea level change out of humanity's control. The Earth, in a broad sense, would be a different planet than the one that existed for the past 10 millennia.

Have we already passed a "tipping point" such that it is now impossible to avoid "dangerous" climate change [*Lovelock*, 2006]? In our estimation, we probably have not, but we must be very close to such a point. It is still feasible to achieve a scenario that keeps additional global warming under 1°C, yielding a degree of climate change that is quantitatively and qualitatively different than under BAU scenarios.

The 'alternative' scenario, designed to keep warming less than 1°C, has a significantly smaller forcing than any of the IPCC scenarios. In recent years net growth of all real world greenhouse gases has run just slightly ahead of the alternative scenario, with the excess due to continued growth of $CO_2$ emissions at about 2%/year. $CO_2$ emissions would need to level out soon and decline before mid-century to approximate the alternative scenario. Moderate changes of emissions growth rate have a marked effect after decades, as shown by comparison to BAU scenarios. Early decreases in emissions growth, albeit small, are the most effective.

The alternative scenario target, keeping added $CO_2$ to ~ 80 ppm between 2000 and 2050, may already be impractical due to the 2%/year growth of $CO_2$ emissions in the past decade. However, the net greenhouse forcing could still meet the alternative scenario target via the combination of a still feasible slowdown and reduction of $CO_2$ emissions together with aggressive absolute reductions of $CH_4$ and $O_3$ and a slowdown in the growth of $N_2O$. Reduction of non-$CO_2$ forcings has benefits for human health and agriculture [*West et al.*, 2005; *Air Pollution Workshop*, www.giss.nasa.gov/meetings/pollution2005], as well as for climate. Reduction of non-$CO_2$ forcings is especially effective in limiting Arctic climate change [section 6.2.1 above; *Shindell et al.*, 2006].

Continued rapid growth of $CO_2$ emissions and infrastructure for another decade may make attainment of the alternative scenario impractical if not impossible. Because wide-scale use of power plants with $CO_2$ sequestration is at least a decade away, near-term emphasis on energy efficiency and renewable energy is needed to minimize construction of previous-generation pulverized-coal power plants. Such power plants are a primary cause of increasing $CO_2$ emissions, and their construction commits the world to high $CO_2$ emissions or high costs of power plant replacement. Potential energy savings from improved efficiency, even in developed countries, are sufficient to cover increased energy demand for the 1-2 decades needed to develop improved energy technologies [*Romm et al.*, 1998; *Pacala and Socolow*, 2004; *NCEP*, 2004].

Human-made aerosols cause a net negative climate forcing that we estimate (Figure 1) as about half the magnitude of positive greenhouse gas forcing. Thus aerosols have diminished global warming in what *Hansen and Lacis* [1990] call a "Faustian bargain". As humanity reduces particulate air pollution in the future, a payment in increased global warming will come due [*Andreae et al.*, 2005]. With appropriate understanding of the aerosol forcing, however, it may be possible to minimize warming effects of air pollution reduction. The direct aerosol forcing can be reduced via emphasis on reduction of black carbon (soot) aerosols that cause warming [*Hansen et al.*, 2000; *Jacobson*, 2001]. The aerosol indirect effect is non-linear, with the first aerosols added having greatest effect [*Menon et al.*, 2002]. Thus retention of a small amount of aerosols, perhaps in remote areas, may minimize warming. Better knowledge of aerosol properties and their effects is needed for analysis.

A scenario that avoids "dangerous" climate change appears to be still technically feasible. The fact that the recent trend of global GHG climate forcing remains close to the 'alternative scenario' that we estimate as necessary to avoid 'dangerous' climate change provides some basis for optimism. In addition, we are entering an era in which the reality of climate change is becoming increasingly apparent, and it is clear that the effects of climate change, especially sea level rise, will be widespread, affecting both developed and developing countries.

Reduction of climate forcing is a global problem, as quantified in Figure 9, and thus it presents an unusual challenge to attain needed international cooperation. For this reason it is particularly important that the scientific understanding of climate change and its implications be transmitted to the public and policy makers, and the message must be understandable and believable if it is to help produce the actions needed to avoid dangerous climate change.

Our conclusion that global temperature is nearing the level of dangerous climate effects implies that little time remains to achieve the international cooperation needed to avoid widespread undesirable consequences. $CO_2$ emissions are the critical issue, because a substantial fraction of these emissions remain in the atmosphere "forever", for practical purposes (Figure 10a). The principal implication is that avoidance of dangerous climate change requires the bulk of coal and unconventional fossil fuel resources to be exploited only under condition that $CO_2$ emissions are captured and sequestered. A second inference is that remaining gas and oil resources must be husbanded, so that their role in critical functions such as mobile fuels can be stretched until acceptable alternatives are available, thus avoiding a need to squeeze such fuels from unconventional and environmentally damaging sources. The task is to achieve a transition to clean carbon-free energy sources, which are essential on the long run, without pushing the climate system beyond a level where disastrous irreversible effects become inevitable.

These stark conclusions about the threat posed by global climate change and implications for fossil fuel use are not yet appreciated by essential governing bodies, as evidenced by ongoing plans to build coal-fired power plants without $CO_2$ capture and sequestration. In our view, there is an acute need for science to inform society about the costs of failure to address global warming, because of a fundamental difference between the threat posed by climate change and most prior global threats. In the nuclear standoff between the Soviet Union and United States, a crisis could be precipitated only by action of one of the parties. In contrast, the present threat to the planet and civilization, with the United States and China now the principal players, requires only inaction in the face of clear scientific evidence of the danger of increased greenhouse gas emissions. Thus scientists are faced with difficult choices between communication of scientific information to the public and focus on basic research, as there are inherent compromises in any specific balance.

**Acknowledgments.** Data that we use for recent greenhouse gas amounts are from the NOAA Earth System Research Laboratory, Global Monitoring Division where we are particularly indebted to Ed Dlugokencky, Steve Montzka, and Jim Elkins for up-to-date data. We thank Ellen Baum, Tom Boden, Curt Covey, Oleg Dubovik, Hans Gilgen, Danny Harvey, Brent Holben, Phil Jones, John Lanzante, Judith Lean, Forrest Mims, Bill Randel, Eric Rignot for data and helpful correspondence, and Darnell Cain for technical assistance. Research support from Hal Harvey of the Hewlett Foundation, Gerry Lenfest, and NASA Earth Science Research Division managers Jack Kaye, Don Anderson, Waleed Abdalati, Phil DeCola, Tsengdar Lee, and Eric Lindstrom is gratefully acknowledged.

**References**

Abdalati, W., W.B. Krabill, E. Frederick, S. Manizade, C. Martin, J. Sonntag, R. Swift, R. Thomas, W. Wright, and J. Yungel (2001), Near-coastalthinning of the Greenland ice sheet, *J. Geophys. Res.*, **106**, 33729-33742.

Andreae, M.O., C.D. Jones, and P.M. Cox (2005) Strong present-day aerosol cooling implies a hot future, *Nature*, **435**, 1187-1190.

Aoki, S., M. Yoritaka, and A. Masuyama (2003), Multidecadal warming of subsurface temperature in the Indian sector of the Southern Ocean, *J. Geophys. Res.*, **108**, C4, 8081, doi:10.1029/2000JC000307.

Archer, D. (2005), Fate of fossil-fuel $CO_2$ in geologic time, *J. Geophys. Res. Oceans*, **110**, doi:10.1029/2004JC002625.

Archer, D. (2006), Methane hydrates and anthropogenic climate change, *Rev. Geophys.* (in press).

Arctic Climate Impact Assessment (ACIA, 2004), *Impacts of a Warming Arctic*, Cambridge Univ. Press, available at www.cambridge.org.

Ausubel, J. (1996), Can technology spare the Earth?, *Amer. Scientist*, **84**, 166-178.

Ausubel, J. (2000), Where is energy going?, *Industrial Physicist*, **6**, 16-19.

Barrett, P.J., C.J. Adams, W.C. McIntosh, C.C. Swisher, and G.S. Wilson (1992) Geochronological evidence supporting Antarctic deglaciation three million years ago, *Nature*, **359**, 816-818.

Bell, G.D., S. Goldenberg, C. Landsea, E. Blake, R. Pasch, M. Chelliah, and K. Mo (2005), Tropical storms, *Bull. Amer. Meteorol. Soc.*, **86**, S26-S29.

Bleck, R. (2002), An oceanic general circulation model framed in hybrid isopycnic-Cartesian Coordinates, *Ocean Modelling*, **4**, 55-88.

Bowen, G.J., T.J. Bralower, M.L. Delaney, G.R. Dickens, D.C. Kelly, P.L. Koch, L.R. Kump, J. Meng, L.C. Sloan, E. Thomas, S.L. Wing, and J.C. Zachos (2006), Eocene hyperthermal event offers insight into greenhouse warming, *EOS Trans. Amer. Geophys. Union*, **87**, 165-169.

British Petroleum (2006), Putting energy in the spotlight, BP Statistical Review of World Energy June 2006, www.bp.com/pdf/statistical_review_of_world_energy_full_report2006.pdf.

Budyko, M.I. (1969), The effect of solar radiation variations on the climate of the earth, *Tellus*, **21**, 611-619.

Cane, M.A., A.C. Clement, A. Kaplan, Y. Kushnir, D. Pozdnyakov, R. Seager, S.E.Zebiak, and R. Murtugudde (1997), Twentieth century sea surface temperature trends, *Science*, **275**, 957-960, 1997.

Collins, W.D., et al. (2006), Radiative forcing by well-mixed greenhouse gases: Estimates from climate models in the IPCC AR4, *J. Geophys. Res.*, 111, D14317, doi: 10.1029/2005JD006713, 2006.

Comiso, J.C. (2002), A rapidly declining perennial sea ice cover in the Arctic, *Geophys. Res. Lett.*, **29(2)**, 1956, doi:10.1029/2002GL015650.

Cox, P.M., R.A. Betts, C.D. Jones, S.A. Spall, and I.J. Totterdell (2000), Acceleration of global warming due to carbon-cycle feedbacks in a coupled climate model, *Nature*, **408**, 184-187.

Crowley, T.J. (1996), Pliocene climates: the nature of the problem, *Mar. Micropaleontol.*, **27**, 3-12.

De Angelis, H., and P. Skvarca (2003), Glacier surge after ice shelf collapse, *Science*, **299**, 1560-1562.

Delworth, T.L., and M.E. Mann (2000), Observed and simulated multidecadal variability in the Northern Hemisphere, *Clim. Dyn.*, **16**, 661-676.

Delworth, T.L., and T.R. Knutson (2000), Simulation of early 20th century global warming, *Science*, **287**, 2246-2250.

Dowsett, H.J., R. Thompson, J. Barron, T. Cronin, F. Fleming, S. Ishman, R. Poore, D. Willard, and T. Holtz (1994), Joint investigations of the Middle Pliocene climate I: PRISM paleoenvironmental reconstructions, *Global Planet. Change*, **9**, 169-195.

Dowsett, H., J. Barron, and R. Poore (1996), Middle Pliocene sea surface temperatures: a global reconstruction, *Mar. Micropaleontol.*, **27**, 13-26.

Droxler, A.W., R.B. Alley, W.R. Howard, R.Z. Poore, and L.H. Burkle (2002), Unique and exceptional marine isotope stage 11: Window to the warm Earth future climate, in *Earth's Climate and Orbital Eccentricity*, A.W. Droxler, R.Z. Poore, and L.H. Burkle (eds.), *Geophys. Mono.* **137**, American Geophysical Union, Washington, DC, pp. 1-14.

Dwyer, G.S., T.M. Cronin, P.A. Baker, M.E. Raymo, J.S. Buzas, and T. Correge (1995), North Atlantic deepwater temperature change during late Pliocene and late Quaternary climatic cycles, *Science*, **270**, 1347-1351.

Emanuel, K. (1987), The dependence of hurricane intensity on climate, *Nature*, **326**, 483-485.

Emanuel, K. (2005), Increasing destructiveness of tropical cyclones over the past 30 years, *Nature*, **436**, 686-688.

Energy Information Administration (2005), Annual Energy Review 2005, www.eia.doe.gov/emeu/aer/contents.html.

Forest, C.E., P.H. Stone and A.P. Sokolov (2006), Estimated PDFs of climate system properties including natural and anthropogenic forcings, *Geophys. Res. Lett.*, **33**, L01705, doi:10.1029/2005GL023977.

Friedlingstein, P., J.L. Dufresne, P.M. Cox, and P. Rayner (2003), How positive is the feedback between climate change and the carbon cycle?, *Tellus*, **55B**, 692-700.

Fung, I.Y., S.C. Doney, K. Lindsay, and J. John (2005), Evolution of carbon sinks in a changing climate, *Proc. Natl. Acad. Sci.*, **102**, 11201-11206.

Gent, P.R., J. Willebrand, T.J. McDougall, and J.C. McWilliams (1995), Parameterizing eddy-induced tracer transports in ocean circulation models, *J. Phys. Oceanogr.*, **25**, 463-474.

Gille, S.T. (2002), Warming of the Southern Ocean since the 1950s, *Science*, **295**, 1275-1277.

Goldenberg, S.B., C.W. Landsea, A.M. Mesta-Nunez, and W.M. Gray (2001), The recent increase in Atlantic hurricane activity: causes and implications, *Science*, **293**, 474-479.

Gray, W. (1984), Atlantic hurricane seasonal frequency. Part I: El Nino and 30 mb quasi-biennial oscillation influences, *Mon. Wea. Rev.*, **112**, 1649-1668.

Gray, W. (2005), *Congressional Record*, testimony to the United States Senate, Disaster Prevention and Prediction Subcommittee, September 20, 2005.

Griffies, S.M. (1998), The Gent-McWilliams skew flux, *J. Phys. Oceanogr.*, **28**, 831-841.

Hansen, J., G. Russell, D. Rind, P. Stone, A. Lacis, S. Lebedeff, R. Ruedy, and L. Travis (1983), Efficient three-dimensional global models for climate studies: Models I and II, *Mon. Weather Rev.*, **111**, 609-662.

Hansen, J., A. Lacis, D. Rind, G. Russell, P. Stone, I. Fung, R. Ruedy, and J. Lerner (1984), Climate sensitivity: Analysis of feedback mechanisms, in *Climate Processes and Climate Sensitivity, Geophys. Monogr. Ser.*, vol. 29, edited by J.E. Hansen and T. Takahashi, pp. 130-163, AGU, Washington, D.C.

Hansen, J.E., and A.A. Lacis (1990), Sun and dust versus greenhouse gases: an assessment of their relative roles in global climate change, *Nature*, **346**, 713-719.

Hansen, J., A. Lacis, R. Ruedy, M. Sato, and H. Wilson (1993), How sensitive is the world's climate?, *Natl. Geogr. Res. Explor.*, **9**, 142-158.

Hansen, J.E., M. Sato, R. Ruedy, A. Lacis, and V. Oinas (2000), Global warming in the twenty-first century: an alternative scenario, *Proc. Natl. Acad. Sci. U.S.A.*, **97**, 9875-9880.

Hansen, J., Defusing the global warming time bomb (2004), *Sci. Amer.*, **290**, 68-77.

Hansen, J., and M. Sato (2004), Greenhouse gas growth rates, *Proc. Natl. Acad. Sci.*, **101**, 16,109-16,114.

Hansen, J. (2005a), A slippery slope: how much global warming constitutes "dangerous anthropogenic interference"?, *Clim.Change*, **68**, 269-279.

Hansen, J. (2005b), Is there still time to avoid 'dangerous anthropogenic interference' with global climate? A tribute to Charles David Keeling, Amer. Geophys. Union, U23D-01, Dec. 6, available www.columbia.edu/~jeh1.

Hansen, J. et al. (2005a), Efficacy of climate forcings, *J. Geophys. Res.*, **110**, D18104, doi:10.1029/2005JD005776.

Hansen, J. et al. (2005b), Earths energy imbalance: confirmation and implications, *Science*, **308**, 1431-1435.

Hansen, J., M. Sato, R. Ruedy, K. Lo, D.W. Lea and M. Medina-Elizade (2006a) Global temperature change, *Proc. Natl. Acad. Sci.*, 103, 14288-14293, doi:10.1073/pnas0606291103.

Hansen, J., et al. (2006b), Climate simulations for 1880-2003 with GISS modelE, to be submitted.

Hawkins, D.G. (2005), Integrated strategies for fossil generation emission reduction, in *Air Pollution as a Climate Forcing: A Second Workshop*, ed. J.E. Hansen, available NASA Goddard Institute for Space Studies, or www.giss.nasa.gov/meetings/pollution2005.

Hays, J.D., J. Imbrie, and N.J. Shackleton (1976), Variations in the Earth's orbit: pacemaker of the ice ages, *Science*, **194**, 1121-1132.

Hearty, P.J., P. Kindler, H. Cheng, and R.L. Edwards (1999), A +20 m middle Pleistocene sea-level highstand (Bermuda and the Bahamas) due to partial collapse of Antarctic ice, *Geology*, **27**, 375-378.

Henderson-Sellers, A., et al. (1998), Tropical cyclones and global climate change: a post-IPCC assessment, *Bull. Amer. Meteorol. Soc.*, **79**, 19-38.

Hoffert, M.I., and C. Covey (1992), Deriving global climate sensitivity from paleoclimate reconstructions, *Nature*, **360**, 573-576.

Hughes, T. (1972), Is the West Antarctic ice sheet disintegrating?, ISCAP Bulletin, no. 1, Ohio State University.

Hughes, T. (1981), The weak underbelly of West Antarctic ice sheet, *J. Glaciology*, **27**, 518-525.

Huybrechts, P., J. Gregory, I. Janssens, and M. Wild (2004), Modelling Antarctic and Greenland volume changes during the 20th and 21st centuries forced by GCM time slice integrations, *Global Planet. Change*, **42**, 83-105.

Intergovernmental Panel on Climate Change (IPCC, 2001), *Climate Change 2001: The Scientific Basis*, eds. Houghton J.T., Ding, Y., Griggs, D.J., Noguer, M., van der Linden, P.J., Dai, X., Maskell, K. & Johnson, C.A, Cambridge Univ. Press, Cambridge, U.K.

Jacobson, M.Z. (2001), Global direct radiative forcing due to multicomponent anthropogenic and natural aerosols, *J. Geophys. Res.*, **106**, 1551-1568.

Johannessen, O.M., and 11 co-authors (2004), Arctic climate change: observed and modeled temperature and sea-ice variability, *Tellus*, **56A**, 328-341.

Jones, C.D., P.M. Cox, and C. Huntingford (2005), Impact of climate-carbon cycle feedbacks on emissions scenarios to achieve stabilization, in

Joos, F., M. Bruno, R. Fink, T.F. Stocker, U. Siegenthaler, C. Le Quere, and J.L. Sarmiento (1996), An efficient and accurate representation of complex oceanic and biospheric models of anthropogenic carbon uptake, *Tellus*, **48B**, 397-417.

Kharecha, P.A., and J. Hansen (2006), in preparation

Kienast, M., T.J.J. Hanebuth, C. Pelejero, and S. Steinke (2003), Synchroneity of meltwater pulse 1a and the Bolling warming: New evidence from the South China Sea, *Geology*, **31**, 67-70.

Koch, D. (2001), Transport and direct radiative forcing of carbonaceous and sulfate aerosols in the GISS GCM, *J. Geophys. Res.*, **106**, 20311-20332.

Kushnir, Y. (1994), Interdecadal variations in North Atlantic sea surface temperature and associated atmospheric conditions, *J. Clim.*, **7**, 141-157.

Kvenvolden, K.A., and T.D. Lorenson (2001), The global occurrence of natural gas hydrate, in *Natural Gas Hydrates – Occurrence, Distribution, and Detection*, *Geophys. Monogr.*, **124**, eds. C.K. Paull and W.P. Dillon, Amer. Geophys. Union, pp. 3-18 (data available at walrus.wr.usgs.gov/globalhydrate/index.html).

Large, W.G., J.C. McWilliams, and S.C. Doney (1994), Oceanic vertical mixing: a review and a model with a nonlocal boundary layer parameterization, *Rev. Geophys.*, **32**, 363-403.

Lean, J., and D. Rind (1998), Climate forcing by changing solar radiation, *J. Climate*, **11**, 3069-3094.

Lean, J. (2000), Evolution of the sun's spectral irradiance since the Maunder Minimum, *Geophys. Res. Lett.*, **27**, 2425-2428.

Lean, J.L., Y.M. Wang, and N.R. Sheeley (2002), The effect of increasing solar activity on the Sun's total and open magnetic flux during multiple cycles: implications for solar forcing of climate, *Geophys. Res. Lett.*, **29** (no. 24), 2224, doi:10.1029/2002GL015880.

Levitus, S., J.I. Anatov, and T.P. Boyer (2005), Warming of the world ocean, 1955-2003, *Geophys. Res. Lett.*, **322**, L02604, doi:10.1029/2004GL021592.

Lindsay, R.W., and J. Zhang (2005), The thinning of Arctic sea ice, 1988-2003: have we passed a tipping point?, *J. Clim.*, in press.

Liu, J., G.A. Schmidt, D.G. Martinson, D. Rind, G.L. Russell, and X. Yuan (2003), Sensitivity of sea ice to physical parameterizations in the GISS global climate model, *J. Geophys. Res.*, **108**, 3053, doi:10.1029/2001JC001167.

Lovelock, James (2006), *The Revenge of Gaia*, 192 pp., Allen Lane, Santa Barbara, CA.

Lyman, J.M., J.K. Willis, and G.C. Gregory (2006), Recent cooling of the upper ocean, *Geophys. Res. Lett.*, **33**, L18604, doi:10.1029/2006GL027033.

Mann, M.E., R.S. Bradley, and M.K. Hughes (1998) Global-scale temperature patterns and climate forcing over the past six centuries, *Nature*, **392**, 779-787.

Mann, M.E., et al. (2003), On past temperatures and anomalous late-20th century warmth, *Eos Trans. Amer. Geophys. Union*, **84**, 256-258.

Mann, M.E., and K.A. Emanuel (2006), Atlantic hurricane trends linked to climate change, *Eos Trans. Amer. Geophys. Union*, **87**, 233-244.

Marland, G., T.A. Boden, and R.J. Andres (2006), Global, regional, and national $CO_2$ emissions, in *Trends: Compendium of Data of Global Change*, Carbon Dioxide Information Analysis Center, Oak Ridge National Laboratory, U.S. Dept. Energy, Oak Ridge, TN.

Matthews, H.D., A. J. Weaver, and K.J. Meissner (2005), Terrestrial carbon cycle dynamics under recent and future climate change, *J. Climate*, **18**, 1609-1628.

Mayfield, M. (2005), *Congressional Record*, testimony to the United States Senate, Disaster Prevention and Prediction Subcommittee, September 20, 2005.

McCulloch, M.T., and T. Esat (2000), The coral record of last interglacial sea levels and sea surface temperatures, *Chem. Geol.*, **169**, 107-129.

Medina-Elizade, M., and D.W. Lea (2005), The mid-Pleistocene transition in the Tropical Pacific, *Science*, **310**, 1009-1012.

Menon, S., A.D. Del Genio, D. Koch, and G. Tselioudis (2002), GCM simulations of the aerosol indirect effect: Sensitivity to cloud parameterization and aerosol burden, *J. Atmos. Sci.*, **59**, 692-713.

Menon, S., and A. Del Genio (2006), Evaluating the impacts of carbonaceous aerosols on clouds and climate, in *Human-Induced Climate Change: An Interdisciplinary Assessment*, eds. Schlesinger et al., Cambridge Univ. Press, in press.

Mercer, J.H. (1978) West Antarctic ice sheet and $CO_2$ greenhouse effect: a threat of disaster, *Nature*, **271**, 321-325.

Miller, R. L., Schmidt, G. A., and Shindell, D. T. (2006), Forced annular variability in the 20th century Intergovernmental Panel on Climate Change Fourth Assessment Report models, *J. Geophys. Res.*, **111**, D18101, doi:10.1029/2005JD006323.

Mudelsee, M. (2001), The phase relations among atmospheric $CO_2$ content, temperature and global ice volume over the past 420 ka, *Quat. Sci. Rev.*, **20**, 583-589.

National Commission on Energy Policy (NCEP, 2004), *Ending the Energy Stalemate: A Bipartisan Strategy to Meet America's Energy Challenges*, 149 pp., www.energycommission.org.

North, G.R. (1984),The small ice cap instability in diffusive climate models, *J. Atmos. Sci.*, **41**, 3390-3395.

Okabe, Y. (2002), Contribution of nuclear power to reduce emission of air pollution from electricity sector in Japan, in *Air Pollution as a Climate Forcing: A Workshop*, ed. J. Hansen, 169 pp., available Goddard Institute for Space Studies, New York.

Otto-Bliesner, B.L., S.J. Marshall, J.T. Overpeck, G.H. Miller, and A. Hu (2006), Simulating Arctic climate warmth and icefield retreat in the last interglaciation, *Science*, **311**, 1751-1753.

O'Niell, B.C., and M. Oppenheimer (2002), Climate change impacts are sensitive to the concentration stabilization path, *Science*, **296**, 1971-1972.

Overpeck, J., K. Hughen, D. Hardy, R. Bradley, R. Case, et al. (1997), Arctic environmental change of the last four centuries, *Science*, **278**, 1251-1256.

Overpeck, J.T., et al. (2005), Arctic system on trajectory to new, seasonally ice-free state, *EOS, Trans. Amer. Geophys. Union*, **86**, 309-312.

Overpeck, J.T., B.L. Otto-Bliesner, G.H. Miller, D.R. Muhs, R.B. Alley, and J.T. Kiehl (2006), Paleoclimate evidence for future ice-sheet instability and rapid sea level rise, *Science*, **311**, 1747-1750.

Pacala, S., and R. Socolow (2004), Stabilization wedges: solving the climate problem for the next 50 years with current technologies, *Science*, **305**, 968-972.

Petit, J.R., J. Jouzel, D. Raynaud, N.I. Barkov, J.M. Barnola, I. Basile, M. Bender, J. Chappellaz, J. Davis, G. Delaygue, and 9 co-authors (1999), 420,000 years of climate and atmospheric history revealed by the Vostok deep Antarctic ice core, *Nature*, **399**, 429-436.

Potter, E.K., T.M. Esat, G. Schellmann, U. Radtke, K. Lambeck, and M.T. Mc-Culloch (2004), Suborbital-period sea-level oscillations during marine isotope substages 5a and 5c, *Earth Planet. Sci. Lett.*, **225**, 191-204.

Rayner, N.A., D.E., Parker, E.B. Horton, C.K. Folland, L.V.Alexander, D.P. Rowell, E.C. Kent, and A. Kaplan (2003), Global analyses of SST, sea ice and night marine air temperatures since the late nineteenth century, *J. Geophys. Res.*, **108**, doi:10.1029/2002JD002670.

Rignot, E.J. (1998), Fast recession of a West Antarctic glacier, *Science*, **281**, 549-551.

Rignot, E., and S.S. Jacobs (2002), Rapid bottom melting widespread near Antarctic ice sheet grounding lines, *Science*, **296**, 2020-2023.

Rignot, E., G. Casassa, P. Gogineni, W. Krabill, A. Rivera, and R. Thomas (2004), Accelerated ice discharge from the Antarctic Peninsula following the collapse of Larsen B ice shelf, *Geophys. Res. Lett.*, **31**, L18401, doi:10.1029/2004GL020697.

Romm, J., M. Levine, M. Brown, and E. Petersen (1998), A roadmap for U.S. carbon reductions, *Science*, **279**, 669-670.

Russell, G.L., J.R. Miller, and L.C. Tsang (1985), Seasonal oceanic heat transports computed from an atmospheric model, *Dynam. Atmos. Oceans*, **9**, 253-271.

Russell, G.L., J.R. Miller, and D.H. Rind (1995), A coupled atmosphere-ocean model for transient climate change, *Atmosphere-Ocean*, **33**, 683-730.

Santer, B.D., et al. (2006), Forced and unforced ocean temperature changes in Atlantic and Pacific tropical cyclogenesis regions, *Proc. Natl. Acad. Sci.* (in press).

Sato, M., J. Hansen, M.P. McCormick, and J.P. Pollack (1993), Stratospheric aerosol optical depths, 1850-1990, *J. Geophys. Res.*, **98**, 22,987-22,994.

Schmidt, G.A., et al. (2006), Present day atmospheric simulations using GISS ModelE: Comparison to in-situ, satellite and reanalysis data. *J. Climate*, **19**,

153-192.

Schneider, S.H., and M.D. Mastrandrea (2005), Probabilistic assessment of "dangerous" climate change and emissions pathways, *Proc. Natl. Acad. Sci.*, **102**, 15728-15735.

Shepherd, A., D.J. Wingham, and J.A.D. Mansley (2002), Inland thinning of the Amundsen Sea sector, West Antarctica, *Geophys. Res. Lett.*, **29(10)**, 1364, doi:10.1029/2001GL014183.

Shepherd, A., D. Wingham, and E. Rignot (2004), Warm ocean is eroding West Antarctic ice sheet, *Geophys. Res. Lett.*, **31(23)**, 402, doi:10.1029/2004GL021106.

Shindell, D., G. Faluvegi, A. Lacis, J. Hansen, R. Ruedy, and E. Aguilar (2006), The role of tropospheric ozone increases in 20$^{th}$ century climate change, *J. Geophys. Res.*, **111**, D08302, doi:10.1029/2005JD006348.

Shine, K.P., J. S. Fuglestvedt, K. Hailemariam, and N. Stuber (2005), Alternatives to the global warming potential for comparing climate impacts of emissions of greenhouse gases, *Clim. Change*, **68**, 281-302.

Siddall, M., E.J. Rohling, A. Almogi-Labin, Ch. Hemleben, D. Meischner, I. Schmelzer, and D.A. Smeed (2003), Sea-level fluctuations during the last glacial cycle, *Nature*, **423**, 853-858.

Sinton, J.E. (2001), Accuracy and reliability of China's energy statistics, *China Econ. Rev.*, **12**, 373-383.

Smith, J., and S. Hitz (2003), Estimating the global impact of climate change, ENV/EPOC/GSP(2003)12, Organization for Economic Co-operation and Development (OECD), Paris.

Sun, S., and R. Bleck (2006), Geographical distribution of the diapycnal component of thermohaline circulations in coupled climate models, *Ocean Modelling*, in press.

Thomas, R.H., W. Abdalati, E. Frederick, W.B. Krabill, S. Manizade, and K. Steffen (2003), Investigation of surface melting and dynamic thinning on Jakobshavn Isbrae, Greenland, *J. Glaciol.*, **49**, 231-239.

Thomas, R., et al. (2004), Accelerated sea-level rise from West Antarctica, *Science*, **306**, 255-258.

Thompson, W.G., and S.L. Goldstein (2005), Open-system coral ages reveal persistent suborbital sea-level cycles, *Science*, **308**, 401-404.

Trenberth, K. (2005) Uncertainty in hurricanes and global warming, *Science*, **308**, 1753-1754.

United Nations (1992), United Nations framework convention on climate change, UNFCCC, www.unfccc.int.

van den Broeke, M. (2005), Strong surface melting preceded collapse of Antarctic Peninsula ice shelf, *Geophys. Res. Lett.*, **32**, L12815, doi:10.1029/2005GL023247.

Walter, K.M., S.A. Zimov, J.P. Chanton, D. Verbyla, and F.S. Chapin (2006), Methane bubbling from Siberian thaw lakes as a positive feedback to climate warming, *Nature*, **443**, 71-75.

Webster, P.J., G.J. Holland, J.A. Curry, and H.R. Chang (2005), Changes in tropical cyclone number, duration, and intensity in a warming environment, *Science*, **309**, 1844-1846.

West, J.J., A.M. Fiore, L.W. Horowitz, and D.L. Mauzerall (2006), Mitigating ozone pollution with methane emission controls: global health benefits, *Proc. Natl. Acad. Sci.*, 103, 3988-3993, doi:10.1073/pnas.0600201103.

Wild, M., P. Calanca, S.C. Scherer, and A. Ohmura (2003), Effects of polar ice sheets on global sea level in high-resolution greenhouse scenarios, *J. Geophys. Res.*, **108**(D5), 4165, doi:10.1029/2002JD002451.

Willis, J.K., Roemmich, D., Cornuelle, B. (2004) Interannual variability in upper ocean heat content, temperature, and thermosteric expansion on global scales, *J. Geophys. Res.*, **109**, C12036, doi:10.1029/2003JC002260.

Yohe, G., N. Andronova, and M. Schlesinger (2004), To hedge or not against an uncertain climate future?, *Science*, **306**, 416-417.

Zimov, S.A., E.A.G. Schuur, and F.S. Chapin (2006), Permafrost and the global carbon budget, *Science*, **312**, 1612-1613.

Zwally, H.J., W. Abdalati, T. Herring, K. Larson, J. Saba, and K. Steffen (2002), Surface melt-induced acceleration of Greenland ice-sheet flow, *Science*, **297**, 218-222.

———

I. Aleinov, S. Bauer, M. Chandler, G. Faluvegi, J. Jonas, P. Kharecha, D. Koch, J. Lerner, L. Nazarenko, Ju. Perlwitz, M. Sato, N. Unger, S. Zhang, Columbia University Earth Institute, New York, NY 10025, USA.

E. Baum, A. Cohen, Clean Air Task Force, Boston, MA USA.

B. Cairns, Ja. Perlwitz, Department of Applied Physics and Applied Mathematics, Columbia University, New York, NY 10025, USA.

V. Canuto, A. Del Genio, T. Hall, J. Hansen, N.Y. Kiang, A. Lacis, R. Miller, D. Rind, A. Romanou, G. Russell, G.A. Schmidt, D. Shindell, NASA Goddard Institute for Space Studies, New York, NY 10025, USA. (jhansen@giss.nasa.gov)

Y. Cheng, K. Lo, V. Oinas, R. Ruedy, R. Schmunk, N. Tausnev, M. Yao, Sigma Space Partners LLC, New York, NY 10025, USA.

E. Fleming, C. Jackman, G. Labow, NASA Goddard Space Flight Center, Greenbelt, MD 20771, USA.

A. Friend, M. Kelley, Laboratoire des Sciences du Climat et de l'Environnement, Orme des Merisiers, Gif-sur-Yvette Cedex, France.

S. Menon, T. Novakov, Lawrence Berkeley National Laboratory, Berkeley, CA 94720, USA.

P. Stone, S. Sun, Massachusetts Institute of Technology, Cambridge, MA 02139, USA.

D. Streets, Argonne National Laboratory, Argonne, IL, USA.

D. Thresher, Department of Earth and Environmental Sciences, Columbia University, New York, NY 10025 USA.